\documentclass[letterpaper,twocolumn,10pt]{article}
\usepackage{usenix}

\usepackage{amsmath}
\usepackage{amssymb}
\usepackage{mathtools}
\usepackage{array}
\usepackage{xspace}
\usepackage{xcolor}
\usepackage{shortcut}
\usepackage{booktabs}
\usepackage{makecell}
\usepackage{graphicx}
\usepackage{multirow}
\usepackage{etoolbox}
\usepackage{enumitem}

\makeatletter
\g@addto@macro\UrlBreaks{\do\-}
\makeatother

\AtBeginEnvironment{equation}{\small}
\AtBeginEnvironment{equation*}{\small}
\AtBeginEnvironment{align}{\small}
\AtBeginEnvironment{align*}{\small}
\AtBeginEnvironment{gather}{\small}
\AtBeginEnvironment{gather*}{\small}
\AtBeginEnvironment{multline}{\small}
\AtBeginEnvironment{multline*}{\small}

\newcolumntype{C}[1]{>{\centering\arraybackslash}p{#1}}

\title{\Large \bf Not an Accident:\\
Exploiting Safety Filter Inadequacy in Humanoid Robots}
\author{
{\large\textnormal{Tiffany Bao}$^{1}$, \textnormal{Shangtao Li}$^{2}$,
\textnormal{Yifan Sun}$^{2}$, \textnormal{Mengwen He}$^{2}$,
\textnormal{Bowei Li}$^{2}$} \\
{\large\textnormal{Alden Wang}$^{2}$, \textnormal{Changliu Liu}$^{2}$,
\textnormal{Ruoyu Wang}$^{1}$, \textnormal{David Brumley}$^{2}$} \\
{\large $^{1}$ Arizona State University; $^{2}$ Carnegie Mellon University}
}
\date{}

\begin{document}

\maketitle

\begin{abstract}

As humanoid robots move from research laboratories toward deployment in human-centered environments such as homes, hospitals, and warehouses, safe operation around nearby people and objects becomes a fundamental requirement.
\Safecontrols have emerged as a promising technique to address this challenge by providing mathematical guarantees that generated control actions do not result in collisions.

This paper shows that these guarantees hold only when the safety parameters are configured correctly.
However, correct parameter configurations are task- and environment-specific, and no general method exists to determine them across diverse deployments.
As a result, \safecontrols shipped with humanoid robots can fail after deployment in specific scenarios.
We show that attackers can exploit these failures through only normal user interactions without any additional privileges or capabilities.
To demonstrate this risk, we present \theattack and develop \tool, an automated tool that identifies vulnerable \insertions for a given task and environment.
Experimental results show that \tool identifies attacks in all vulnerable scenarios in the SPARK dataset.
Compared with a baseline approach, \tool discovers three times as many attacks.
We also show that \tool is able to generate attacks against a real humanoid and cause it to collide while executing tasks. 
These results demonstrate that current \safecontrols introduce exploitable vulnerabilities that can cause safety violations, which reveals that existing safety mechanisms can themselves become a source of safety risk for humanoid robots.

\end{abstract}

\section{Introduction}
As humanoid robots move into human-centered environments such as homes and hospitals, ensuring safe operation around people and objects becomes a fundamental requirement that one must meet, because even a single collision can cause physical injury, property damage, or disruption of critical tasks~\cite{surgery}.

\Safecontrols have emerged as a promising approach to this challenge.
They monitor the control generated by a robot's nominal controller and intervene whenever the nominal control violates safety constraints.
Optimization-based \safecontrols compute an alternative control by solving a constrained optimization problem based on formulations such as safety indices~\cite{ssa,zhao2023sis}, safe sets~\cite{ssa,pssa}, and control barrier functions (CBFs)~\cite{cbf,ames2019cbfsurvey}.
By construction, the resulting control satisfies the specified safety constraints whenever the optimization problem is feasible.
Thanks to these guarantees, \safecontrols have attracted significant industrial interest and have entered pilot deployments in large industrial sectors.

However, existing \safecontrols can still fail to prevent collisions.
One well-known limitation is \infeasibility~\cite{agrawal2021iccbf,chen2021backupcbf}, where the underlying optimization problem admits no feasible solution and therefore no safe control exists.

We identify a second limitation that has received little attention.
Existing \safecontrols rely on safety-related parameters whose correct values depend on both the task and the environment.
Improper parameter configurations can cause the controller to generate controls that satisfy the optimization problem but nevertheless fail to maintain safety.
Humanoid robots exacerbate both limitations, as their high-dimensional articulated bodies and operation in cluttered, dynamic environments make optimization infeasibility more likely~\cite{khazoom2022humanoidcbf,peng2025biped} and accurate parameter configuration significantly more difficult~\cite{yun2025siadaptation}.

Worse, we discover that existing optimization-based \safecontrols only optimize the current control step.
They neither predict future states nor account for future optimization feasibility.
As a result, a controller may successfully resolve the current safety violation while simultaneously steering the robot toward a future state where no feasible recovery exists.

In this paper, we show that these limitations expose a previously unrecognized attack surface.
An adversary can exploit them to steer a humanoid robot into a \emph{safety-inadequate} state where the safety controller can no longer prevent collision.
Unlike prior attacks that compromise software, spoof sensors, or manipulate low-level control signals, our attack requires only legitimate task requests submitted through the robot's normal user interface.
By carefully selecting intermediate goals or task locations, an attacker can intentionally drive the robot toward a safety-inadequate state that its \safecontrol cannot avoid.

We argue that this vulnerability has a real-world implication.
Consider a bedside assistant robot deployed in a hospital.
While the robot delivers medication to a patient, another patient or visitor requests that it also retrieve a glass of water or pick up medication from a nearby room.
Because the request appears legitimate and authorized, the task planner naturally inserts it into the execution schedule.
The synthesized intermediate goal then steers the robot toward a safety-inadequate state.
Depending on the robot's fail-safe strategy, the robot may fail the original task, collide with nearby people or medical equipment, or even fall during recovery~\cite{emergencystop}.
Since every interaction is legitimate, such attacks are difficult to detect or attribute.

To assess the practicality and impact of this attack, we develop \tool, short for \emph{\textbf{S}afe-control \textbf{I}nadequacy via \textbf{R}eachable-goal \textbf{E}xploitation \textbf{N}exus}.
\tool is an automated framework that identifies safety-inadequate states in a given \emph{scenario}, which is a combination of an environment and a target task, and synthesizes adversarial task goals that steer the robot into those states.
We evaluate \tool against seven representative optimization-based \safecontrol algorithms, including Control Barrier Functions (CBFs)~\cite{cbf}, Safe Set Algorithm (SSA)~\cite{ssa}, Sublevel Safe Set (SSS)~\cite{wei2019energy}, and their variants~\cite{pssa}.
The evaluation demonstrates successful attacks against every evaluated \safecontrol in both simulation and on a Unitree G1 humanoid robot.
Compared with a random-search baseline, the Cross-Entropy Method (CEM) used by \tool discovers \moretimes times of attacks in \datasize vulnerable scenarios from SPARK, a popular testbed that includes a benchmark for \safecontrol evaluation~\cite{spark}.

In summary, this paper makes the following contributions:

\begin{itemize}[itemsep=2pt]
    \item We identify \emph{safe-control inadequacy} as a previously unrecognized security vulnerability in optimization-based humanoid \safecontrols and show that attackers can exploit the vulnerability to cause safety issues such as collisions.
To the best of our knowledge, this is the first security attack that targets optimization-based \safecontrols in humanoid robots.
    \item We present \theattack and develop \tool, an automated framework that discovers safety-inadequate robot states and synthesizes adversarial task goals through normal user interactions.
    \item We evaluate \tool against seven representative \safecontrol algorithms in simulation and on a Unitree G1 humanoid robot. The results show successful attacks against every evaluated \safecontrol, and that \tool discovers \moretimes times as many attacks as identified by the random-search baseline.
\end{itemize}

\section{Background and Related Work}
\label{sec:background}

\subsection{Robotic Security}
\label{sec:robotic_security}

Prior work on robotic security has shown that robots expose attack surfaces across multiple layers of the software and hardware stack.
Existing work identifies vulnerabilities in network services, firmware, communication middleware, sensors, actuators, and application interfaces~\cite{mayoral2018rsf,yaacoub2021robotics}, while attacks such as industrial-robot ransomware demonstrate that software compromise can translate directly into unsafe physical behavior~\cite{mayoral2019akerbeltz}.
These risks have been studied across a wide range of robotic platforms.
In autonomous vehicles, researchers have demonstrated attacks against in-vehicle networks and remote interfaces~\cite{koscher2010car,checkoway2011car}, localization through sensor-fusion and GPS manipulation~\cite{shen2020drift}, and camera and LiDAR perception pipelines~\cite{zhang2025,cao2019lidar,sun2020lidar,cao2021invisible}.
In robotic vehicles, prior work has used fuzzing and simulation to identify policy, maneuver, and supply chain vulnerabilities~\cite{kim2021pgfuzz,muller2022drivetruth,song2023acero,kim2023patchverif}, and multi-robot systems are vulnerable to semantic attacks against collaborative navigation~\cite{yeke2025raven}.

Humanoid robots introduce additional security challenges.
Contrary to autonomous vehicles and conventional mobile robots, humanoids possess articulated bodies with high-dimensional state spaces, configuration-dependent collision geometry, and state-dependent control authority.
These properties create attack opportunities beyond conventional software, communication, and perception vulnerabilities.
Recent work has identified security issues in commercial humanoids, including authentication vulnerabilities in Bluetooth~\cite{cveble}, Wi-Fi~\cite{cvewifi}, and DDS-based communication~\cite{cvedds}, as well as privacy and telemetry risks in commodity humanoid platforms~\cite{mayoral2025humanoid,mayoral2025humanoidattackvector}.
However, little attention has focused on vulnerabilities in humanoid-specific control and mechanisms such as humanoid safety algorithms.

\subsection{Robot Dynamics}

Following common practice in nonlinear control~\cite{khalil2002nonlinear}, we model a humanoid as a robot system governed by control-affine dynamics.
Let $\mathbf{x} \in \mathcal{X} \subseteq \mathbb{R}^{N_x}$ denote the \emph{state} of the robot, which is a high-dimensional vector that describes information relevant to the robot at a particular time, where $\mathcal{X}$ is the set of admissible states and $N_x$ is the dimension of the state.
Depending on the robot and control abstraction, the state may include quantities such as joint positions, joint velocities, and the position and orientation of the robot base.
Similarly, let $\mathbf{u} \in \mathcal{U} \subseteq \mathbb{R}^{N_u}$ denote the \emph{control input}, or \emph{action}, which specifies commands that affect the robot's motion, such as desired joint velocities or accelerations, where $\mathcal{U}$ is the set of admissible controls and $N_u$ is the dimension of the control input.

The evolution of the state under a control input is described by the control-affine dynamics
\begin{equation}
\dot{\mathbf{x}}
=
\mathbf{f}(\mathbf{x})
+
\mathbf{g}(\mathbf{x})\mathbf{u},
\qquad
\mathbf{u} \in \mathcal{U},
\label{eq:dynamics}
\end{equation}
where $\mathbf{f}(\mathbf{x})$ describes the uncontrolled evolution of the system and $\mathbf{g}(\mathbf{x})$ characterizes how the control input affects the state. For example, under a first-order dynamics model, the state may consist of joint positions and $\mathbf{u}$ specifies joint velocities; under a second-order model, the state additionally contains joint velocities and $\mathbf{u}$ specifies joint accelerations.

Modern robotic systems typically employ a \emph{nominal controller} or policy to generate actions for accomplishing a task.
We denote its output as the \emph{reference control} $\mathbf{u}_{\mathrm{ref}}$.
The reference control represents the action that the robot would execute in the absence of an additional safety intervention.
It may be produced by a conventional controller, such as a proportional-integral-derivative (PID) controller~\cite{astrom1995pid} or inverse kinematics~\cite{siciliano2009robotics}, or by a learning-based policy~\cite{levine2016e2e}.
In increasingly autonomous robotic systems, the task policy may also incorporate foundation models, e.g., vision-language-action (VLA) models~\cite{brohan2023rt2,kim2024openvla,black2024pi0}, to interpret high-level observations and commands and ultimately produce or condition the robot's reference actions.
The reference controller is primarily responsible for task performance and need not itself enforce the safety constraints considered below.

\subsection{Safety Filter Algorithms}

A \emph{\safecontrol} operates between the nominal policy and the robot actuators.
Rather than replacing the nominal controller, it receives $\mathbf{u}_{\mathrm{ref}}$ and produces a safe control $\mathbf{u}_{\mathrm{safe}}$ that closely resembles the reference action while satisfying safety requirements.
This architecture separates task execution from safety enforcement and allows the same safety mechanism to operate on reference actions generated by different upstream controllers.

A broad family of model-based safety filters characterize safety through a scalar named \emph{safety index} denoted by $\phi(\mathbf{x})$.
Following the convention used in the work in the robotics safety literature~\cite{ssa,wei2019energy}, $\phi(\mathbf{x}) \leq 0$ represents the desired safe region, whereas $\phi(\mathbf{x}) > 0$ indicates a safety violation.
For $M$ simultaneous safety constraints, the desired safe set is
\begin{equation}
\mathcal{X}_{s}
=
\left\{
\mathbf{x}\in\mathcal{X}
\mid
\phi^{[i]}(\mathbf{x})\leq0,;
\forall i\in[M]
\right\}.
\label{eq:safe_set}
\end{equation}

For collision avoidance, for example, a basic safety specification can be constructed from the distance $d(\mathbf{x},\mathcal{O})$ between the robot and an obstacle $\mathcal{O}$ and a required minimum separation $d_{\min}$.

At a high level, optimization-based \safecontrols solve
\begin{equation}
\begin{aligned}
\mathbf{u}_{\mathrm{safe}}
=
\arg\min_{\mathbf{u}} \quad
& \left\|
\mathbf{u}-\mathbf{u}_{\mathrm{ref}}
\right\|_{Q_{\mathbf{u}}}^{2} \\
\text{subject to} \quad
& \dot{\mathbf{x}}
=
\mathbf{f}(\mathbf{x})
+
\mathbf{g}(\mathbf{x})\mathbf{u}, \\
& \dot{\phi}^{[i]}(\mathbf{x},\mathbf{u})
\leq
\rho^{[i]}(\phi^{[i]}(\mathbf{x})),
\quad \forall i \in [M],\\
& \mathbf{u}\in\mathcal{U},
\end{aligned}
\label{eq:general_safe_filter}
\end{equation}
where $Q_{\mathbf{u}}$ weights deviations from the reference action and $\rho(\cdot)$ specifies the safety algorithm's constraint on the evolution of the safety index.

Note that a \safecontrol does not necessarily modify the reference control.
When the safety index is below zero, e.g., the humanoid maintains a distance greater than $d_{\min}$ from obstacles, the \safecontrol preserves the reference control, such that $\mathbf{u}_{\mathrm{safe}} = \mathbf{u}_{\mathrm{ref}}$.
The \safecontrol modifies the reference control only when necessary to enforce the safety constraint.
In this paper, we refer to a \safecontrol as being \emph{engaged} when it actively modifies the reference control.

Intuitively, once the robot approaches or enters an unsafe region, the \safecontrol constrains how quickly the robot goes back to safe region, which, analytically, is how quickly $\phi$ may drop.
We refer to this as the \emph{escape rate}: the required rate at which the controller drives the system away from a safety violation and toward the safe region.
Different safety-filter families impose different escape rates, and their corresponding hyperparameters directly influence both the aggressiveness of safety intervention and the deviation from $\mathbf{u}_{\mathrm{ref}}$.

\paragraph{Safe Set Algorithm (SSA)}
The Safe Set Algorithm (SSA) activates corrective constraints when a safety index reaches or exceeds the boundary of the safe set~\cite{ssa,zhao2023sis}. For multiple constraints, its optimization can be written as
\begin{equation}
\begin{aligned}
\min_{\mathbf{u}} \quad
& \left\|
\mathbf{u}-\mathbf{u}_{\mathrm{ref}}
\right\|_{Q_{\mathbf{u}}}^{2} \\
\text{subject to} \quad
& \dot{\mathbf{x}}
=
\mathbf{f}(\mathbf{x})
+
\mathbf{g}(\mathbf{x})\mathbf{u},\\
& \dot{\phi}^{[i]}(\mathbf{x},\mathbf{u})
\leq -\eta_{\mathrm{ssa}},
&& \text{if } \phi^{[i]}(\mathbf{x})>0,\\
& \dot{\phi}^{[i]}(\mathbf{x},\mathbf{u})
\leq 0,
&& \text{if } \phi^{[i]}(\mathbf{x})=0,
\end{aligned}
\label{eq:ssa}
\end{equation}
for all $i\in[M]$, where $\eta_{\mathrm{ssa}}>0$ controls the required decrease of the safety index after a violation. Thus, $\eta_{\mathrm{ssa}}$ directly specifies an escape-rate requirement: increasing $\eta_{\mathrm{ssa}}$ requires the system to reduce a positive safety index more rapidly and generally results in a stronger correction to the reference control.

\paragraph{Sublevel Safe Set Algorithm (SSS)} A related method, the Sublevel Safe Set (SSS) algorithm, instead imposes
\begin{equation}
\dot{\phi}^{[i]}(\mathbf{x},\mathbf{u})
\leq
-\lambda_{\mathrm{sss}}\phi^{[i]}(\mathbf{x}),
\qquad
\text{if } \phi^{[i]}(\mathbf{x})\geq0.
\label{eq:sss}
\end{equation}
where $\lambda_{\mathrm{sss}}>0$ scales the required decrease of the safety index with the current magnitude of that index.
Unlike the constant escape rate of SSA, SSS scales the required correction with the magnitude of the safety violation.

\paragraph{Control Barrier Function (CBF)}
Control Barrier Function (CBF) filters continuously constrain the evolution of the safety index~\cite{cbf,ames2019cbfsurvey}. Using a linear class-$\mathcal{K}$ function, the constraint takes the form
\begin{equation}
\dot{\phi}^{[i]}(\mathbf{x},\mathbf{u})
\leq
-\lambda_{\mathrm{cbf}}
\phi^{[i]}(\mathbf{x}),
\qquad
\forall i\in[M],
\label{eq:cbf}
\end{equation}
where $\lambda_{\mathrm{cbf}}>0$ controls the allowable evolution of $\phi$. The resulting safe action is obtained by minimizing the deviation from $\mathbf{u}_{\mathrm{ref}}$ subject to Eq.~\ref{eq:cbf}. In contrast to SSA and SSS, the CBF constraint remains active even when $\phi<0$. Consequently, $\lambda_{\mathrm{cbf}}$ affects not only how the controller responds to violations but also how much freedom the nominal controller retains while the system is already inside the safe set. CBFs are related to barrier certificates~\cite{prajna2004barrier} and have been analyzed for robustness under disturbances~\cite{xu2015robustcbf} and extended to discrete-time systems for bipedal robots~\cite{agrawal2017dcbf}.

\paragraph{Slack-Relaxed Filters (R-SSA, R-SSS, and R-CBF)}
The relaxed variants of SSA, SSS, and CBF introduce one non-negative slack variable $s_i$ for each safety constraint and optimize the control and slack variables jointly. Their common formulation is
\begin{equation}
\begin{aligned}
\min_{\mathbf{u},\mathbf{s}} \quad
& \|\mathbf{u}-\mathbf{u}_{\mathrm{ref}}\|_{Q_{\mathbf{u}}}^{2}
+ w_s\|\mathbf{s}\|_{p}^{p} \\
\text{subject to} \quad
& \dot{\mathbf{x}}
=
\mathbf{f}(\mathbf{x})
+
\mathbf{g}(\mathbf{x})\mathbf{u},\\
& \dot{\phi}^{[i]}(\mathbf{x},\mathbf{u})
\leq
\rho_A\!\left(\phi^{[i]}(\mathbf{x})\right)+s_i,
\quad \forall i\in\mathcal{I}_A,\\
& \mathbf{u}\in\mathcal{U},\\
& \mathbf{s}\geq0,
\end{aligned}
\label{eq:relaxed_filters}
\end{equation}
where $\|\mathbf{s}\|_{p}$ is the $\ell_p$ norm of the slack vector, $A\in\{\mathrm{R\text{-}SSA},\mathrm{R\text{-}SSS},\mathrm{R\text{-}CBF}\}$ identifies the relaxed filter, and
\begin{equation}
\rho_A(\phi)
=
\begin{cases}
-\eta_{\mathrm{rssa}}, & A=\mathrm{R\text{-}SSA},\\
-\lambda_{\mathrm{rsss}}\phi, & A=\mathrm{R\text{-}SSS},\\
-\lambda_{\mathrm{rcbf}}\phi, & A=\mathrm{R\text{-}CBF}.
\end{cases}
\label{eq:relaxed_filter_rates}
\end{equation}
As in their non-relaxed counterparts, R-SSA and R-SSS invoke the optimization when at least one enabled safety index is positive; during that optimization, $\mathcal{I}_A$ contains the enabled indices with $\phi^{[i]}(\mathbf{x})\geq0$. R-CBF instead invokes the optimization continuously, and $\mathcal{I}_A$ contains every enabled safety index, including those with $\phi^{[i]}(\mathbf{x})<0$.

The optimizer yields $\mathbf{u}_{\mathrm{safe}}$ and $\mathbf{s}^{*}$. Each $s_i^{*}$ is the amount by which the corresponding derivative bound is violated: $s_i^{*}=0$ recovers the original filter constraint, while $s_i^{*}>0$ weakens it. Consequently, provided that the control set $\mathcal{U}$ is nonempty, the slack variables prevent conflicting safety constraints from making the optimization infeasible. The SPARK implementation of these filters uses $p=2$ by default and also supports $p=1$. The slack weight $w_s>0$ determines a direct tradeoff within the single optimization: a larger value discourages relaxation and may require a greater deviation from $\mathbf{u}_{\mathrm{ref}}$, whereas a smaller value permits more safety-constraint violation in order to follow the reference control more closely.

\paragraph{Projected Safe Set Algorithm (p-SSA)}
For humanoid robots, simultaneously enforcing many safety constraints can make the optimization in SSA infeasible, i.e., there may be no control input $\mathbf{u}\in\mathcal{U}$ that satisfies all required escape rates. The Projected Safe Set Algorithm (p-SSA) also introduces slack, but unlike the single-stage relaxed filters above, it first finds the minimum relaxation of the infeasible safety constraints and then computes the safe control.

Specifically, p-SSA operates in two phases. In the first phase, it computes the minimum relaxation required to obtain a feasible constraint set:
\begin{equation}
\begin{aligned}
\left(\mathbf{u}^{\dagger},\mathbf{s}^{*}\right)
=
\arg\min_{\mathbf{u},\mathbf{s}} \quad
& \|\mathbf{s}\|_{p,Q_s}^{p} \\
\text{subject to} \quad
& \dot{\mathbf{x}}
=
\mathbf{f}(\mathbf{x})
+
\mathbf{g}(\mathbf{x})\mathbf{u},\\
& \dot{\phi}^{[i]}(\mathbf{x},\mathbf{u})
\leq
-\eta_{\mathrm{pssa}} + s_i,\\
& \qquad \text{if } \phi^{[i]}(\mathbf{x})\geq0,\ \forall i\in[M],\\
& \mathbf{u}\in\mathcal{U},\\
& \mathbf{s}\geq0.
\end{aligned}
\label{eq:pssa_projection}
\end{equation}
Here, $s_i$ represents the amount by which the escape-rate requirement of the $i$-th safety constraint must be relaxed. When all constraints can be simultaneously satisfied, $\mathbf{s}^{*}=\mathbf{0}$ and no relaxation is necessary.

The second phase of p-SSA fixes the minimum relaxation $\mathbf{s}^{*}$ and computes the control that most closely follows the reference control:
\begin{equation}
\begin{aligned}
\mathbf{u}_{\mathrm{safe}}
=
\arg\min_{\mathbf{u}} \quad
& \|\mathbf{u}-\mathbf{u}_{\mathrm{ref}}\|_{Q_{\mathbf{u}}}^{2} \\
\text{subject to} \quad
& \dot{\mathbf{x}}
=
\mathbf{f}(\mathbf{x})
+
\mathbf{g}(\mathbf{x})\mathbf{u},\\
& \dot{\phi}^{[i]}(\mathbf{x},\mathbf{u})
\leq
-\eta_{\mathrm{pssa}} + s_i^{*},\\
& \qquad \text{if } \phi^{[i]}(\mathbf{x})\geq0,\ \forall i\in[M],\\
& \mathbf{u}\in\mathcal{U}.
\end{aligned}
\label{eq:pssa_control}
\end{equation}
Thus, p-SSA follows the same constant escape-rate principle as SSA, parameterized by $\eta_{\mathrm{pssa}}$, but permits the requested escape rate to be minimally relaxed when multiple safety requirements cannot be simultaneously satisfied. By separating constraint relaxation from reference-control tracking, p-SSA prioritizes minimizing safety-constraint violations before minimizing deviation from $\mathbf{u}_{\mathrm{ref}}$.

\section{Attack Scenario and Threat Model}

\paragraph{Attack Scenario}
We consider an attack scenario where a humanoid robot operates in an environment containing obstacles and serves multiple users by executing user-specified tasks.
For example, a service humanoid may deliver dishes to customers in a restaurant or distribute medication to patients in a healthcare facility.
An attacker can be one of the users served by the humanoid and interacts with the robot through the same task interface available to benign users.

The robots may receive multiple tasks over time and dynamically schedule their execution according to priorities, deadlines, or execution costs.
Consequently, the scheduler may insert a newly submitted task into the existing task sequence.
For example, a delivery robot may prioritize a newly submitted request that is geographically close to its current location before resuming an earlier delivery.

We show that this benign task-insertion functionality can be exploited to induce safety-related failures.
Consider an original task that requires the humanoid to move from a starting state toward a target goal.
The attacker's objective is to identify an intermediate goal, which we call an \emph{insertion}, such that the original task succeeds when executed directly but fails after the robot first accomplishes the inserted goal and resumes its original task.
In other words, although both the original goal and the inserted goal are individually legitimate, their composition drives the robot into a state from which its \safecontrol prevents successful completion of the original task.

We assume that a task-insertion mechanism exists in the robot's application layer; the specific mechanism through which an attacker inserts a new task is outside the scope of this work.

\paragraph{Non-Attack Scenario}
A related but distinct scenario is one in which the task itself directly causes the humanoid to fail; for example, the humanoid collides with an obstacle while executing the task, regardless of whether an insertion has occurred.
This scenario is outside the scope of our work because the failure is directly attributable to the task itself and can therefore be readily detected.

\paragraph{Threat Model}
We consider a knowledgeable but unprivileged attacker, who knows the \safecontrol deployed on the target humanoid, including its relevant configuration parameters.
This assumption is realistic when the target platform is commercially available: an attacker may obtain an identical robot and analyze its software and firmware offline~\cite{unitreefirmware, jailbreak}.
Robot software and firmware may also be recovered from public resources~\cite{unitreeonline}, hardware debugging interfaces such as JTAG, or software vulnerabilities~\cite{jailbreak}.
The attacker can then reverse engineer the control stack to recover the safety-filter implementation and its parameters.

We additionally assume that the attacker knows the relevant environment geometry, including the locations of obstacles.
This information may be available because the attack takes place in a public or otherwise observable physical environment.
The attacker also knows the target task, including its starting and ending locations, which may be obtained through reconnaissance.
Given the knowledge of the robot, environment, and the target task, the attacker can construct an offline simulation or physical replica where candidate insertions can be evaluated before launching the attack in the real world.

However, at attack time, the attacker's capabilities are substantially restricted.
The attacker interacts with the humanoid only through the legitimate high-level task interface that is available to ordinary users.
Attackers cannot exploit vulnerabilities to achieve, e.g., remote code execution, privileged software access, arbitrary control of joint-level commands, modification of the \safecontrol, sensor spoofing, communication hijacking, or physical tampering with the target robot.

\section{Goal Injection Attacks}
\label{sec:goal_injection}

In this section, we present the \insertionattack.
We first state the safety failures existing in the current \safecontrol algorithms.
Then we present how attackers can exploit the safety failures maliciously, so that they can trap the victim humanoids to collisions.

\subsection{\SafeControls Failures}
\label{sec:attack_surface}

In this subsection, we show that the safety requirements imposed by a
\safecontrol can lead to two distinct failure modes.
When the configured escape rate is too small, the \safecontrol may not drive the robot away
from an obstacle sufficiently quickly.
Conversely, when the configured
escape rate is too large, the required safety response may exceed the
robot's available control authority, making the safe-control optimization
infeasible.
A goal injection attack exploits these limitations by steering
the humanoid into a state where one of these failure modes can be triggered.

Recall from Equation~\ref{eq:general_safe_filter} that an
optimization-based \safecontrol computes $\mathbf{u}_{\mathrm{safe}}$ such that 
$\dot{\phi}^{[i]}(\mathbf{x},\mathbf{u}) \leq \rho^{[i]}(\phi^{[i]}(\mathbf{x})), \forall i\in[M]$.
In the rest of the analysis, we focus on one constraint $i$ without the loss of generality.
For the ease of presentation, we let $\phi\equiv\phi^{[i]}$, $\rho\equiv\rho^{[i]}$, and $s \equiv s_i$.
We then define the corresponding
\emph{required escape rate}, denoted by $\rate$, as
\begin{equation}
\rate =
-\rho(\phi(\mathbf{x})), \quad \rate > 0
\label{eq:demand}
\end{equation}

For SSA-family and CBF/SSS-family filters, $\rate=\eta$ and $\rate=\lambda\phi(\mathbf{x})$, respectively.
For filters with slack variables, $\rate=\eta + s$ or $\rate=\lambda\phi(\mathbf{x}) + s$, where $\eta$, $\lambda$, and $s$ are generic shorthand for the corresponding family-specific
configuration parameters.
Thus, despite their different formulations, these filters constrain the rate at which
the safety index is allowed to evolve.
Combining the simplified notation, for the rest of this section, the optimization problem that \safecontrols seek to compute is the following equation:

\begin{equation}
\begin{aligned}
\mathbf{u}_{\mathrm{safe}}
=
\arg\min_{\mathbf{u}} \quad
& \| \mathbf{u}-\mathbf{u}_{\mathrm{ref}} \|_{Q_{\mathbf{u}}}^{2} \\
\text{subject to} \quad
& \dot{\mathbf{x}}
=
\mathbf{f}(\mathbf{x})
+
\mathbf{g}(\mathbf{x})\mathbf{u},\\
& \dot{\phi}(\mathbf{x},\mathbf{u}) \leq -\rate,\\ & \mathbf{u}\in\mathcal{U}.
\end{aligned}
\end{equation}

\paragraph{Insufficient Escape Rate.}
The first failure mode occurs when the escape rate imposed by the
\safecontrol is too small relative to the state of the humanoid when
the \safecontrol engages.
Consider a collision-specific safety index with a \emph{relative degree} $r$, following the definition of existing \safecontrols~\cite{ssa, pssa, nguyen2016ecbf, xiao2019hrdcbf}: 

\begin{equation}
\small
  \begin{aligned}
\phi(\mathbf{x})
=
\phi_0(\mathbf{x})
    + k_1\dot{\phi}_0(\mathbf{x})
    + k_2\ddot{\phi}_0(\mathbf{x})
+ \cdots
    + k_{r-1}\phi_0^{(r-1)}(\mathbf{x}),\\
\text{where}\quad 
\phi_0(\mathbf{x})
=
d_{\min}-d(\mathbf{x}),
  \end{aligned}
\end{equation}

where the subscript in $\phi_0$ identifies the underlying physical
specification and $d(\mathbf{x})$ denotes the clearance between a robot
collision volume and an obstacle.
In the following, $f^{(j)}$ denotes
the $j$-th total time derivative of any function $f(\cdot)$.
Because $d_{\min}$ is constant, we have 
\begin{equation}
\small
\phi(\mathbf{x})
=
d_{\min}
-
d(\mathbf{x})
-
k_1\dot{d}(\mathbf{x})
-
k_2\ddot{d}(\mathbf{x})
-
\cdots
-
k_{r-1}d^{(r-1)}(\mathbf{x})
\label{eq:distance_safety_index}
\end{equation}

\noindent And differentiating Equation~\ref{eq:distance_safety_index} gives
\begin{equation}
\dot{\phi}(\mathbf{x})
=
-
\bigl(
\dot{d}(\mathbf{x})
+
k_1\ddot{d}(\mathbf{x})
+
k_2\dddot{d}(\mathbf{x})
+
\cdots
+
k_{r-1}d^{(r)}(\mathbf{x})
\bigr).
\label{eq:phi_dot_distance}
\end{equation}

By Equation~\ref{eq:demand},
$\rate>0$ represents the minimum rate at which
the filter requires the safety index to decrease.
Combining Equation~\ref{eq:phi_dot_distance}
and~\ref{eq:demand} gives
\begin{equation}
\dot{d}(\mathbf{x})
+
k_1\ddot{d}(\mathbf{x})
+
k_2\dddot{d}(\mathbf{x})
+
\cdots
+
k_{r-1}d^{(r)}(\mathbf{x})
\geq
\rate.
\label{eq:general_distance_escape_constraint}
\end{equation}

Let $\actual$ be the actual value of the left hand side.
The evolution of the distance satisfies
\begin{equation}
\dot{d}(\mathbf{x})
+
k_1\ddot{d}(\mathbf{x})
+
k_2\dddot{d}(\mathbf{x})
+
\cdots
+
k_{r-1}d^{(r)}(\mathbf{x})
= \actual \geq \rate.
\label{eq:clearance_ode_general}
\end{equation}

Equation~\ref{eq:clearance_ode_general} exposes an important limitation
of high related degree \safecontrols: when $r>1$, it does not
require $\dot{d}(\mathbf{x})\geq0$. The clearance $d(\mathbf{x})$ may therefore continue to
decrease \emph{after} the \safecontrol engages, even while the filter
successfully drives $\phi(\mathbf{x})$ downward.

Therefore, there exist engagement states with positive clearance but sufficiently adverse approach dynamics for which the inherited motion consumes the remaining clearance before the safety response can reverse it.
At some future time $t_c$, the robot can then satisfy
Equation~\ref{eq:clearance_ode_general} while
$d(\mathbf{x}(t_c))<0$.
Decreasing the safety index does not by itself certify that the robot avoids physical contact.

In details, we provide a close-form expression of the clearance function $d(\mathbf{x})$ via solving the ordinary differential equation, and discuss its possibility of being negative in Appendix~\ref{app:fixed_rate_clearance}.

Note that this failure does not imply that a \emph{small}, instead of \emph{insufficient}, escape rate is inherently unsafe.
Rather, it arises from the combination of an escape rate and an unfavorable engagement state.
If the \safecontrol engages early, the available clearance may be sufficient for even a relatively small escape rate.
A goal injection attack changes this condition by steering the humanoid into a state with unfavorable approach dynamics before the critical safety constraint becomes active.
Once the original task resumes, the filter may engage from a state for which its configured escape rate is insufficient to recover the clearance before collision.

\paragraph{Insufficient Control Authority.}
Increasing the escape rate does not eliminate the attack surface.
A second failure mode occurs when the requested escape rate exceeds what
the robot can physically realize at its current state.

For the control-affine dynamics in Equation~\ref{eq:dynamics}, the
time derivative of a safety index is
\begin{equation}
\begin{aligned}
\dot{\phi}(\mathbf{x})
&=
\frac{\partial\phi}{\partial\mathbf{x}}
\mathbf{f}(\mathbf{x})
+
\frac{\partial\phi}{\partial\mathbf{x}}
\mathbf{g}(\mathbf{x})\mathbf{u}\\
&=
L_f\phi(\mathbf{x})
+
L_g\phi(\mathbf{x})\mathbf{u},
\end{aligned}
\label{eq:lie_derivative_attack}
\end{equation}
where $L_f\phi(\mathbf{x})$ captures the evolution of the safety index
due to the uncontrolled system dynamics and $L_g\phi(\mathbf{x})$
captures its sensitivity to the control input.

We define the \emph{control authority} with respect to $\phi$ as the
largest decrease in the safety index that can be produced by an
admissible control:
\begin{equation}
C_{\phi}(\mathbf{x})
=
\max_{\mathbf{u}\in\mathcal{U}}
\bigl(
-L_g\phi(\mathbf{x})\mathbf{u}
\bigr).
\label{eq:control_authority}
\end{equation}

For a symmetric box-constrained control set
\begin{equation}
\mathcal{U}
=
\{
\mathbf{u}
\mid
|u_j|\leq\bar{u}_j,
\quad
j=1,\ldots,N_u
\},
\end{equation}
where $N_u$ is the number of control inputs. The control authority then
has the closed form
\begin{equation}
C_{\phi}(\mathbf{x})
=
\sum_{j=1}^{N_u}
\bar{u}_j
\left|
[L_g\phi(\mathbf{x})]_j
\right|.
\label{eq:control_authority_closed}
\end{equation}

As defined in Equation~\ref{eq:demand},
$\rate$ is nonnegative whenever the
corresponding constraint requires the safety index to decrease.

A control satisfying the safety requirement exists if and only if
\begin{equation}
\min_{\mathbf{u}\in\mathcal{U}}
\bigl(
L_f\phi(\mathbf{x})
+
L_g\phi(\mathbf{x})\mathbf{u}
\bigr)
\leq
-\rate.
\end{equation}
By the definition of $C_{\phi}$, this condition is equivalent to
\begin{equation}
L_f\phi(\mathbf{x})
-
C_{\phi}(\mathbf{x})
\leq
-\rate,
\end{equation}
or
\begin{equation}
C_{\phi}(\mathbf{x})
-
\rate
-
L_f\phi(\mathbf{x})
\geq
0.
\label{eq:feasibility_condition}
\end{equation}

We therefore define the \emph{feasibility margin}
\begin{equation}
\gamma(\mathbf{x})
=
C_{\phi}(\mathbf{x})
-
\rate
-
L_f\phi(\mathbf{x}).
\label{eq:feasibility_margin}
\end{equation}
A single safety constraint is feasible if and only if
\begin{equation}
\gamma(\mathbf{x})\geq0.
\end{equation}
When
\begin{equation}
\gamma(\mathbf{x})<0,
\end{equation}
no admissible control can achieve the escape rate requested by the
\safecontrol.

Importantly, $\mathbf{u}_{\mathrm{ref}}$ does not appear in
Equation~\ref{eq:feasibility_margin}. The reference control affects which
feasible control is selected by the minimum-deviation objective above,
but it does not determine whether the safety constraints themselves are
feasible. Feasibility is instead a property of the current robot state,
environment, safety specification, and control limits.

For multiple simultaneously active safety constraints, individual
feasibility is necessary but not sufficient for joint feasibility.
Restoring the constraint superscript, the authority, required rate, and
margin associated with $\phi^{[i]}$ satisfy
\begin{equation}
\gamma^{[i]}(\mathbf{x})
=
C_{\phi}^{[i]}(\mathbf{x})
-
\rate^{[i]}
-
L_f\phi^{[i]}(\mathbf{x})
\geq0,
\qquad
\forall i\in[M],
\end{equation}
but different constraints may require mutually incompatible controls.
The exact joint-feasibility condition is
\begin{equation}
\begin{aligned}
\exists\mathbf{u}\in\mathcal{U}, \quad
L_g\phi^{[i]}(\mathbf{x})\mathbf{u}
&\leq
-\rate^{[i]} -
L_f\phi^{[i]}(\mathbf{x}),
\qquad \forall i\in[M].
\end{aligned}
\label{eq:joint_feasibility}
\end{equation}

\paragraph{Lack of Proactive Failure Avoidance}
Existing \safecontrol algorithms operate myopically.
At each control cycle, they compute only the control input for the next time step and intervene only after the safety index exceeds a predefined threshold.
Consequently, the computed safe control $\mathbf{u}_{safe}$ may drive the humanoid into a state from which the safety controller can no longer recover.
Because the controller neither predicts future states nor accounts for the long-term consequences of its actions, it cannot proactively prevent such failures.

In summary, the two failures expose a fundamental tradeoff in the configuration of \safecontrol:
if the escape rate is too small, the controller may recover too slowly from a dangerous higher-order state, and if the escape rate is too large, the required control input may exceed the robot's available control authority, which renders the safety optimization infeasible.
Furthermore, both the recovery requirements and the available control authority depend on the robot configuration and the surrounding environment.
As a result, a single fixed parameter configuration cannot provide adequate safety guarantees across all states that an articulated humanoid may encounter.
Combined with the lack of proactive failure avoidance, the current design provides no mechanism to eliminate either failure.

\section{Exploiting Safety Failures Adversarially}
\label{sec:exploiting_safety_failures}

The two failure modes described above are both \emph{state dependent}.
Whether the escape rate is sufficient depends on the state at which the
\safecontrol engages, while whether the safety optimization is feasible
depends on the control authority available at the current state.
Consequently, there exist robot states from which a \safecontrol can fail
even though the same filter operates correctly from other states.

This observation suggests a direct attack strategy: identify a
failure-inducing state $\mathbf{x}^{*}$ and place the humanoid in that
state before it executes a target task. In practice, however, directly
searching for and setting such a state is difficult for two reasons.
First, the state space of a humanoid is high dimensional. As described in
Section~\ref{sec:background}, $\mathbf{x}$ may contain the positions and
velocities of many joints in addition to the robot base state. A
humanoid with $N$ controllable degrees of freedom can therefore have a
state space with dimension on the order of $N$ under first-order
dynamics and $2N$ under second-order dynamics. Exhaustively searching
this space for failure-inducing states quickly becomes intractable.

This challenge is fundamentally different from that of a rigid mobile
system. Consider, for example, an autonomous vehicle approaching a
stationary obstacle along a fixed direction. Its collision behavior can
often be characterized using a small number of state variables, such as
the distance to the obstacle and the vehicle's longitudinal velocity.
An adversary searching for a dangerous state could therefore reason
directly about a critical approach speed or stopping distance.

A humanoid does not admit the same reduction. Its geometry changes with
its joint configuration, and the available control authority depends on
the configuration-dependent kinematics of individual links. The same
wrist position, for example, can be realized by multiple arm
configurations with different joint positions, velocities, and
Jacobians, and therefore different safety responses. Moreover, several
robot links may simultaneously interact with different obstacles.
Thus, a failure-inducing humanoid state cannot generally be described
by a single quantity such as approach speed. Searching directly over
$\mathbf{x}$ requires reasoning about a high-dimensional articulated
configuration.

Second, under our threat model, an attacker does not have an interface
for directly assigning $\mathbf{x}$. The attacker cannot set individual
joint positions or velocities, inject arbitrary low-level controls, or
modify the internal state of the robot. The only available interface is
the same high-level task interface exposed to benign users.

Although the attacker cannot directly manipulate the robot state
$\mathbf{x}$, the attacker can indirectly influence it through
high-level reaching goals. Specifically, executing a goal $\mathbf{G}$
from an initial state $\mathbf{x}_0$ induces a resulting robot state
$\mathbf{x}_{\mathbf{G}}$. We denote this closed-loop state transition as
\begin{equation}
\mathcal{T}(\mathbf{x}_0,\mathbf{G})
=
\mathbf{x}_{\mathbf{G}}.
\label{eq:goal_state_mapping}
\end{equation}
The mapping $\mathcal{T}$ captures the execution of the nominal
controller together with the deployed \safecontrol and need not be
available in closed form.

This observation reduces the attack from searching directly over the
high-dimensional robot state space,
\begin{equation}
\mathbf{x}^{*}\in\mathcal{X},
\end{equation}
to searching over legitimate task goals,
\begin{equation}
\mathbf{G}^{*}\in\mathcal{G},
\end{equation}
where $\mathcal{G}$ denotes the set of goals accepted by the robot's
task interface. For the reaching tasks considered in this work,
$\mathbf{G}$ is a Cartesian target and therefore has substantially lower
dimension than the full humanoid state $\mathbf{x}$.

The attacker can thus use the robot's own controller to realize the
desired state transition: rather than directly specifying a
failure-inducing joint-level state, the attacker searches for a
high-level goal whose execution places the humanoid in a state from
which the subsequent task triggers one of the safety failures identified
in Section~\ref{sec:attack_surface}.

\subsection{Attack Formalization}
\label{sec:definition}

Consider an original task that moves the humanoid from the state induced
by an initial goal $\mathbf{G}_0$ to a destination $\mathbf{G}_1$.  The
attacker inserts an intermediate goal $\mathbf{G}'$, changing the task
sequence from $\mathbf{G}_0\!\rightarrow\!\mathbf{G}_1$ to
$\mathbf{G}_0\!\rightarrow\!\mathbf{G}'\!\rightarrow\!\mathbf{G}_1$.
Let $\mathsf{Succ}(\mathbf{x},\mathbf{G})\in\{0,1\}$ indicate whether
executing goal $\mathbf{G}$ from state $\mathbf{x}$ succeeds, and let
$\mathcal{T}(\mathbf{x},\mathbf{G})$ denote the terminal state produced
by that execution.

\paragraph{\InsertionAttack.} The \insertionattack problem is to find an
inserted goal $\mathbf{G}'\in\mathcal{G}$ such that
\begin{equation}
\begin{aligned}
\mathsf{Succ}(\mathbf{x}_0,\mathbf{G}_1) &= 1,\\
\mathsf{Succ}(\mathbf{x}_0,\mathbf{G}') &= 1,\\
\mathsf{Succ}(
    \mathcal{T}(\mathbf{x}_0,\mathbf{G}'),
    \mathbf{G}_1
) &= 0.
\end{aligned}
\label{eq:goal_insertion_conditions}
\end{equation}

The first condition establishes that the target task is normally
successful. The second prevents the attacker from trivially choosing an
inherently invalid or unreachable intermediate goal. The third captures
the attack itself: executing the legitimate intermediate goal changes
the humanoid state such that subsequently executing the original goal
fails.

Crucially, $\mathbf{G}'$ need not itself be close to an obstacle or
directly cause a safety violation. Its purpose is to induce a particular
robot state $\mathbf{x}_{\mathrm{ins}}$. Depending on the failure
mechanism, this state may cause the robot to approach an obstacle with
unfavorable higher-order dynamics, resulting in an insufficient escape
rate, or may place the articulated body in a configuration with
insufficient control authority, resulting in safe-control infeasibility.
The safety failure can therefore occur only during the subsequent
$\mathbf{G}'\rightarrow\mathbf{G}_1$ leg, even though
$\mathbf{G}_0\rightarrow\mathbf{G}'$ is itself safe.

This reduction is important from both an attack and a security
perspective. Instead of requiring privileged access to manipulate a
high-dimensional internal robot state, the attacker searches a
low-dimensional space of authorized task inputs and lets the robot's own
nominal controller realize the corresponding state transition. The
resulting trajectory is generated by legitimate robot software, the
state observed by the \safecontrol remains truthful, and the injected
goal is itself a valid command. The adversarial effect arises solely
from selecting a goal whose induced state makes a subsequent legitimate
task encounter a safety-filter failure.

\subsection{Finding Inserting Goals}
\label{sec:finding_inserting_goals}

The formalization above defines when an inserted goal constitutes a
successful attack, but it does not directly yield a closed-form objective
over the Cartesian goal space.  The attacker can evaluate candidates in a
simulator with the same nominal controller, robot model, obstacle geometry,
and \safecontrol $\psi$.  The attacker's task is therefore to search for an
inserted goal $\mathbf{G}'$ such that the original task succeeds, the
inserted goal is itself executable, but the two-stage execution
$[\mathbf{G}', \mathbf{G}_1]$ causes \safecontrol to fail.

Exhaustive search is inefficient for finding such attacks.
Each candidate $\mathbf{G}'$ must be evaluated through closed-loop execution:
the collision state induced by reaching $\mathbf{G}'$ depends on inverse
kinematics, robot dynamics, the nominal controller, and the known
\safecontrol.
Inspired by existing work in search-guided fuzzing~\cite{Chen2018Hawkeye, Bohme2017AFLGo, Lyu2019MOPT,Chen2018Angora,fioraldi2020afl++}, we use the Cross-Entropy Method (CEM)~\cite{rubinstein1999cem} to focus simulation trials on regions of the admissible goal space that are more likely to produce failure-inducing collision states, with a self-defined score for guidance.

For a candidate inserted goal $\mathbf{G}'$, the search first checks that
the inserted goal is legitimate:
\begin{equation}
\mathsf{Succ}_{\psi}(\mathbf{x}_0,\mathbf{G}') = 1.
\label{eq:whitebox_insertion_gate}
\end{equation}
Only candidates that pass this gate are evaluated as attacks.  We score a
candidate by simulating the full attacked task under $\psi$:
\begin{equation}
S(\mathbf{G}')
=
s_{\psi}\!\left(
  \mathsf{Exec}_{\psi}
  \bigl(
    \mathcal{T}_{\psi}(\mathbf{x}_0,\mathbf{G}'),
    \mathbf{G}_1
  \bigr)
\right).
\label{eq:whitebox_cem_objective}
\end{equation}
The score is the sum of a discrete outcome bonus and a continuous guidance
term:
\begin{equation}
s_{\psi}(r)
=
b(\ell(r)) - d_{\mathrm{collision}}(r),
\label{eq:whitebox_score_definition}
\end{equation}
where $r$ is the simulation run result returned by $\mathsf{Sim}_{\psi}$,
$\ell(r)$ is the simulation outcome label,
$d_{\mathrm{collision}}(r)$ is the obstacle clearance at the collision from the
inserted goal back to the original goal.  The outcome
bonus is
\begin{equation}
b(\ell)=
\begin{cases}
2, & \ell=\textsc{Collision},\\
1, & \ell=\textsc{Timeout},\\
0, & \ell=\textsc{Reached}.
\end{cases}
\label{eq:whitebox_outcome_bonus}
\end{equation}
Thus, confirmed failures dominate near misses, while the continuous terms
rank candidates within the same outcome class.  In particular, smaller
collision clearance increases the score, guiding CEM toward inserted goals
that return control to the original task near the obstacle.

\begin{figure}[t]
\small
\begin{tabular}{@{}r p{0.86\linewidth}@{}}
\toprule
\multicolumn{2}{@{}p{\linewidth}@{}}{\textbf{Input:}
initial state $\mathbf{x}_0$, original goal $\mathbf{G}_1$,
goal space $\mathcal{G}$, filter $\psi$, simulator $\mathsf{Sim}_{\psi}$,
budget $B$, batch size $m$, elite fraction $\alpha$} \\
\multicolumn{2}{@{}l}{\textbf{Output:}
ranked inserted goals and confirmed attacks} \\
\midrule
1  & $(\mu,\sigma) \leftarrow \mathsf{InitGaussian}(\mathcal{G})$ \\
2  & $\mathcal{R} \leftarrow \emptyset$;\quad $n \leftarrow 0$ \\
3  & \textbf{while} $n < B$ \textbf{do} \\
4  & \quad $\mathcal{B} \leftarrow
      \mathsf{Clip}_{\mathcal{G}}\!\left(
      \{\mathbf{g}_i \sim \mathcal{N}(\mu,\mathrm{diag}(\sigma^2))\}_{i=1}^{m}
      \right)$ \\
5  & \quad $\mathcal{S} \leftarrow \emptyset$ \\
6  & \quad \textbf{for each} $\mathbf{G}' \in \mathcal{B}$ \textbf{do} \\
7  & \quad\quad \textbf{if} $\neg\mathsf{Admissible}(\mathbf{G}')$
      \textbf{ then continue} \\
8  & \quad\quad $r_{\mathrm{ins}} \leftarrow
      \mathsf{Sim}_{\psi}(\mathbf{x}_0,[\mathbf{G}'])$ \\
9  & \quad\quad \textbf{if} $\neg\mathsf{Succ}(r_{\mathrm{ins}})$
      \textbf{ then continue} \\
10 & \quad\quad $r_{\mathrm{atk}} \leftarrow
      \mathsf{Sim}_{\psi}(\mathbf{x}_0,[\mathbf{G}',\mathbf{G}_1])$ \\
11 & \quad\quad $score \leftarrow s_{\psi}(r_{\mathrm{atk}})$ \\
12 & \quad\quad $attack \leftarrow \mathsf{Fail}(r_{\mathrm{atk}})$ \\
13 & \quad\quad $\mathcal{S} \leftarrow
      \mathcal{S} \cup \{(\mathbf{G}', score, attack)\}$ \\
14 & \quad\quad $\mathcal{R} \leftarrow
      \mathcal{R} \cup \{(\mathbf{G}', score, attack)\}$ \\
15 & \quad\quad $n \leftarrow n + 1$ \\
16 & \quad \textbf{end for} \\
17 & \quad $\mathcal{E} \leftarrow
      \mathsf{Top}_{\lceil\alpha|\mathcal{S}|\rceil}(\mathcal{S})$ \\
18 & \quad $(\mu,\sigma) \leftarrow
      \mathsf{FitGaussian}(\{\mathbf{G}' : (\mathbf{G}',\cdot,\cdot)\in\mathcal{E}\})$ \\
19 & \textbf{end while} \\
20 & \textbf{return} $\mathsf{SortByScore}(\mathcal{R})$ \\
\bottomrule
\end{tabular}
\caption{\tool's CEM-based search algorithm.}
\label{fig:cem_insertion_search}
\end{figure}

Figure~\ref{fig:cem_insertion_search} separates attack validity from search
guidance.  Lines 6--9 enforce the legitimacy of the inserted goal: the
attacker may submit only goals that the robot can execute as standalone
tasks.  Lines 10--12 then evaluate the actual attack by executing the
inserted goal followed by the original goal under \safecontrol.
CEM uses the resulting score to update its Gaussian search
distribution in Lines 17--18, concentrating later samples around inserted
goals that either already cause failure or move the system closer to a
failure-inducing collision state.

\section{Implementation}

We develop \tool as a Python module in the SPARK v1.0 simulation framework~\cite{spark}.
The core implementation consists of approximately 1,400 lines of code, and the complete codebase with the adaptation layer and experiment analysis comprises approximately 6,800 lines of code.

\section{Evaluation}

Our evaluation aims to answer the following questions:
\smallskip
\begin{itemize}[nosep]
  \item \textbf{Vulnerability Presence}: Does \theattack exist for each \safecontrol under concrete deployment scenarios (Section~\ref{sec:rq1})?
  \item \textbf{Attack Realism}: Can \theattack compromise real humanoid robots and cause safety violations such as collisions (Section~\ref{sec:rq2})?
  \item \textbf{Attack Effectiveness}: How effectively can \tool discover \theattack across different \safecontrols and deployment scenarios (Section~\ref{sec:rq3})?
\end{itemize}

We build our evaluation on SPARK~\cite{spark}, a humanoid robotics testbed that provides simulation environments, a diverse set of \safecontrol implementations, and a benchmark suite for evaluating them.
The benchmark covers a broad range of scenarios, including static and dynamic obstacles, full-body and upper-body tasks, and different safety index formulations.
This diversity enables a comprehensive evaluation of both \safecontrols and \theattack.
All experiments are conducted on macOS running on an Apple M3 processor with 128GB of memory.

\subsection{\SafeControl Vulnerability Presence}
\label{sec:rq1}

Although our analysis (Section~\ref{sec:goal_injection}) shows that \safecontrol algorithms can be vulnerable, it remains unclear whether \emph{concrete} scenarios exist in which an attacker can trigger these vulnerabilities.
One possible approach is to directly use the SPARK benchmark.
However, using the benchmark as-is makes it difficult to isolate and reason about the feasibility of \theattack in a concrete environment.
Some SPARK tasks may fail even without an attack, while successful attacks in complex benchmark environments may involve many interacting obstacles and constraints, making it difficult to attribute the observed failure specifically to \theattack.
We therefore study simplified scenarios to establish the existence and feasibility of \theattack under controlled conditions.

In this experiment, we employ a simple enumeration approach to identify attacks in an environment containing a single obstacle.
We sample tasks from the SPARK benchmark, where each task specifies a starting point and an ending point that the humanoid must reach.
For each sampled task, we search jointly for an \insertion and an obstacle location such that the \safecontrol fails during the attacked execution, resulting in either a collision or a deadlock.
To ensure that the observed failure is caused by \theattack rather than an inherently infeasible task, we compare task execution with and without the attack.
We consider an attack valid only if the original task succeeds without \theattack, while the same task fails after \theattack is introduced.

To construct a feasible \theattack, we enumerate candidate \insertions within a preconfigured $30 \times 30 \times 30$~cm$^3$ three-dimensional region, sampling one point every 5~cm.
For each candidate \insertion, we search for an obstacle location satisfying three conditions, following the definition of \theattack (Section~\ref{sec:definition}): (1) the original task (from the starting point to the ending point) succeeds; (2) the first leg of the attacked execution (from the starting point to the \insertion) succeeds; and (3) the second leg (from the \insertion to the original ending point) fails.

To efficiently identify candidate obstacle locations, we compute the swept volumes of the humanoid along the three trajectories: the original trajectory, the first leg of the attacked trajectory, and the second leg.
We select obstacle locations that lie outside the swept volumes of both the original trajectory and the first leg, but inside the swept volume of the second leg.
Among these locations, we prioritize those farther from the swept volumes of the original trajectory and the first leg.
Intuitively, this procedure places the obstacle where it is unlikely to interfere with either the original task or the first leg, while positioning it to interact with the humanoid during the second leg and potentially trigger a \safecontrol failure.
Finally, we validate each candidate by simulating the complete attacked execution from the starting point to the \insertion and subsequently to the original ending point.

Note that this enumeration procedure is sound but not complete: every reported attack is validated through simulation, but the procedure may miss \insertions that would also produce successful attacks.
Our objective in this experiment is to establish whether concrete instances of \theattack exist for the evaluated \safecontrols, rather than to exhaustively discover all possible attacks in a given environment.
We study the problem of efficiently identifying attacks in Section~\ref{sec:rq3}.

 \begin{table}[t]
\centering
\small
\caption{Validated goal-insertion attacks found by enumeration in single-obstacle environments.}
\smallskip
\label{tab:rq1-main-results}
\begin{tabular}{@{}cccc@{}}
\toprule
\textbf{\SafeControl} & \textbf{\makecell{Valid\\Attacks}} &
  \textbf{\makecell{Min Original\\Distance (cm)}} &
  \textbf{\makecell{Initial Collision\\Distance (cm)}} \\
\midrule
SSA   & \checkmark & 0.0187 & -0.0152 \\
R-SSA & \checkmark & 0.0241 & -0.0298 \\
p-SSA  & \checkmark & 0.0858 & -0.0443 \\
CBF   & \checkmark & 0.2729 & -0.0181 \\
R-CBF & \checkmark & 0.0841 & -0.0273 \\
R-SSS & \checkmark & 0.0625 & -0.0458 \\
SSS   & \checkmark & 0.0187 & -0.0102 \\
\bottomrule
\end{tabular}
\end{table}

As shown in Table~\ref{tab:rq1-main-results}, the results demonstrate that \theattack is feasible against all evaluated \safecontrols.
Across the evaluated SPARK tasks, we are able to identify an \insertion and an obstacle placement for which the original task succeeds, the first leg of the attacked task succeeds, and the \safecontrol is subsequently engaged but fails during the second leg.
These results provide concrete instances of the vulnerabilities established analytically in Section~\ref{sec:attack_surface}, demonstrating that they can be reached through goal insertion rather than requiring direct manipulation of the humanoid's internal state.

\begin{figure}[t]
\centering
\includegraphics[width=0.8\columnwidth]{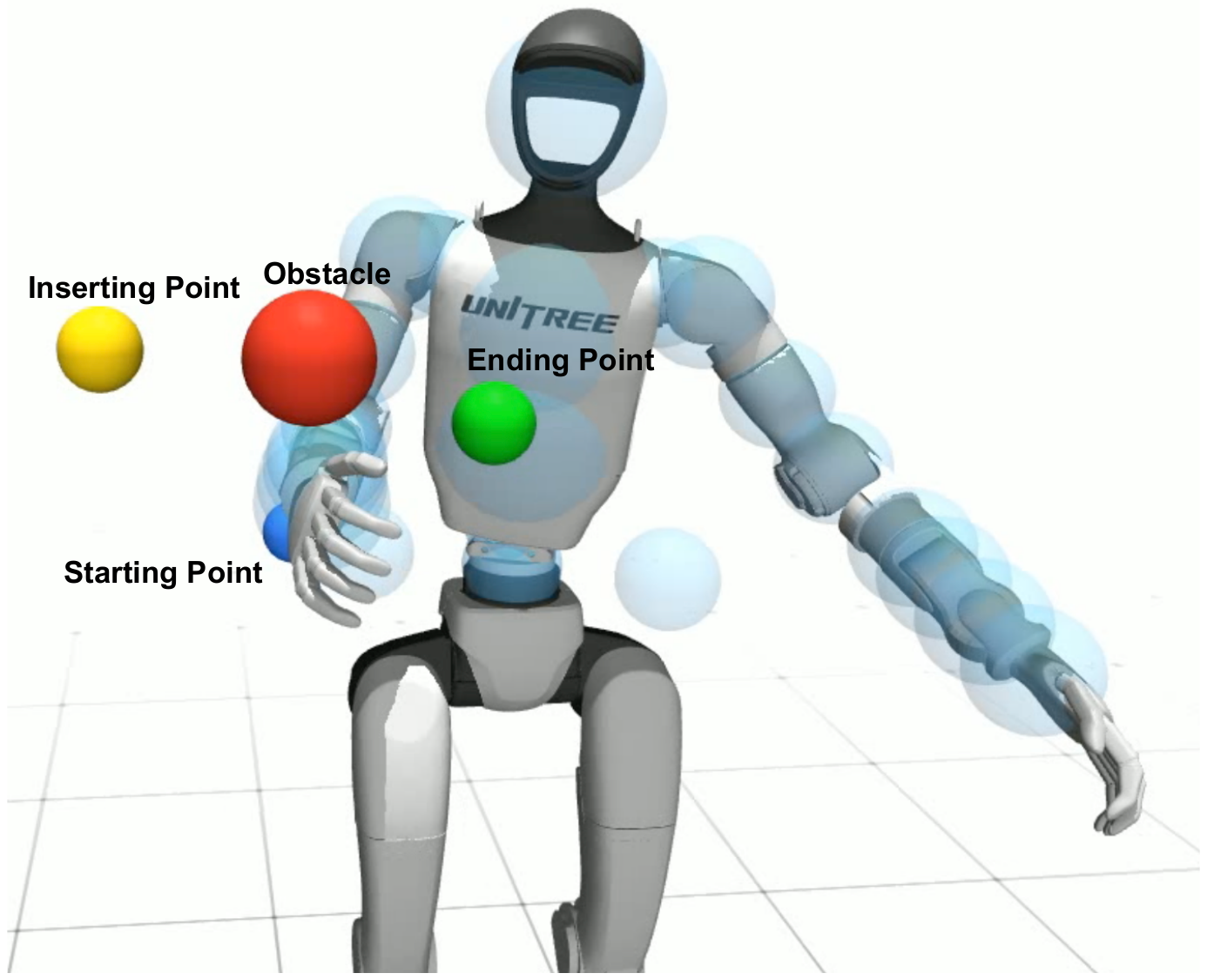}
\caption{A goal insertion attack against SSA.}
\label{fig:casestudy}
\end{figure}

\paragraph{Case Study}
To understand how an attack defeat the \safecontrol, we study one successful case.
In this case, the obstacle is posed in front of the chest humanoid, and the original task is to move the right arm from a point below the obstacle to a point on the robot's left side to the obstacle while move the left arm to a fixed point and stay hold (default in SPARK), as shown in Figure~\ref{fig:casestudy}.
We identify an \insertion against the Safe Set Algorithm (SSA), where we insert a middle target right to the obstacle.
With this inserted goal, a collision happens when the right arm moves from the middle point to the ending point.   

In more details, SSA was engaged before the collision happen.
However, the safety escaping rate, $\eta$, associated with the \safecontrol is configured to be too slow compared to the current arm speed of the humanoid.
By introducing the insertion point, we manage to induce the humanoid's arm to a speed that beyond the SSA's configured safety escaping rate.
Therefore, although SSA was triggered when the distance between the humanoid arm and the obstacle is below threshold, and SSA successfully adjusts the control to move away from the obstacle, given the current speed of the humanoid and its inertia, the SSA-adjusted control was not able to turn the arm away in time. 

\begin{figure}[t]
\centering
\includegraphics[width=0.9\columnwidth]{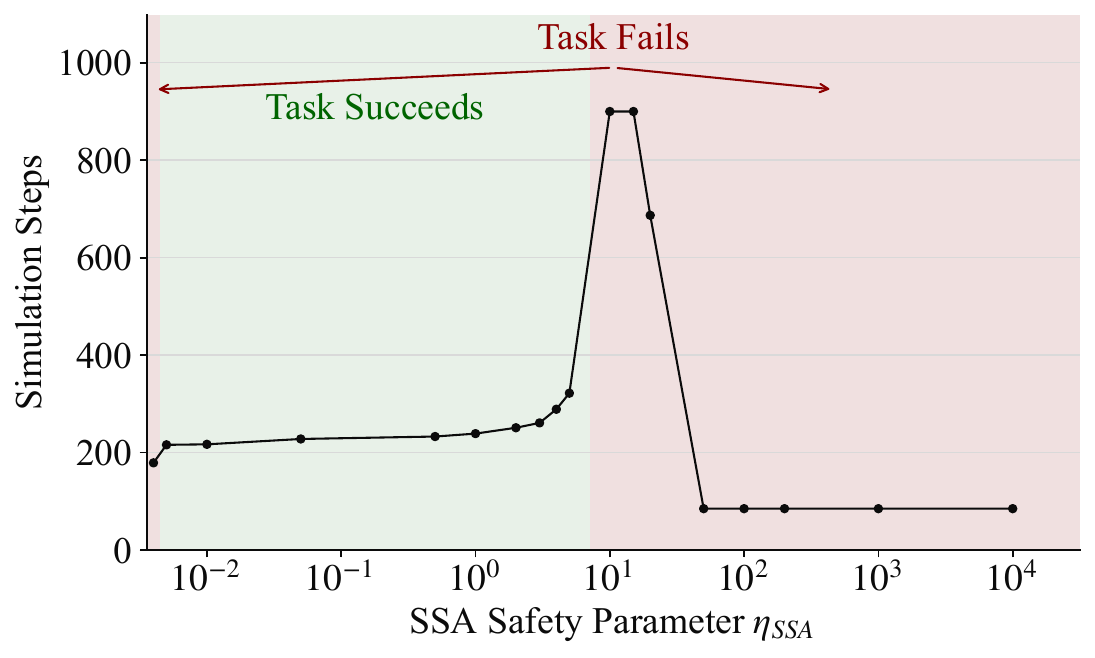}
  \caption{Original task success over SSA safety escaping rate.}
\label{fig:eta-steps-task}
\end{figure}

\begin{figure*}[t]
\centering
\includegraphics[width=0.7\textwidth]{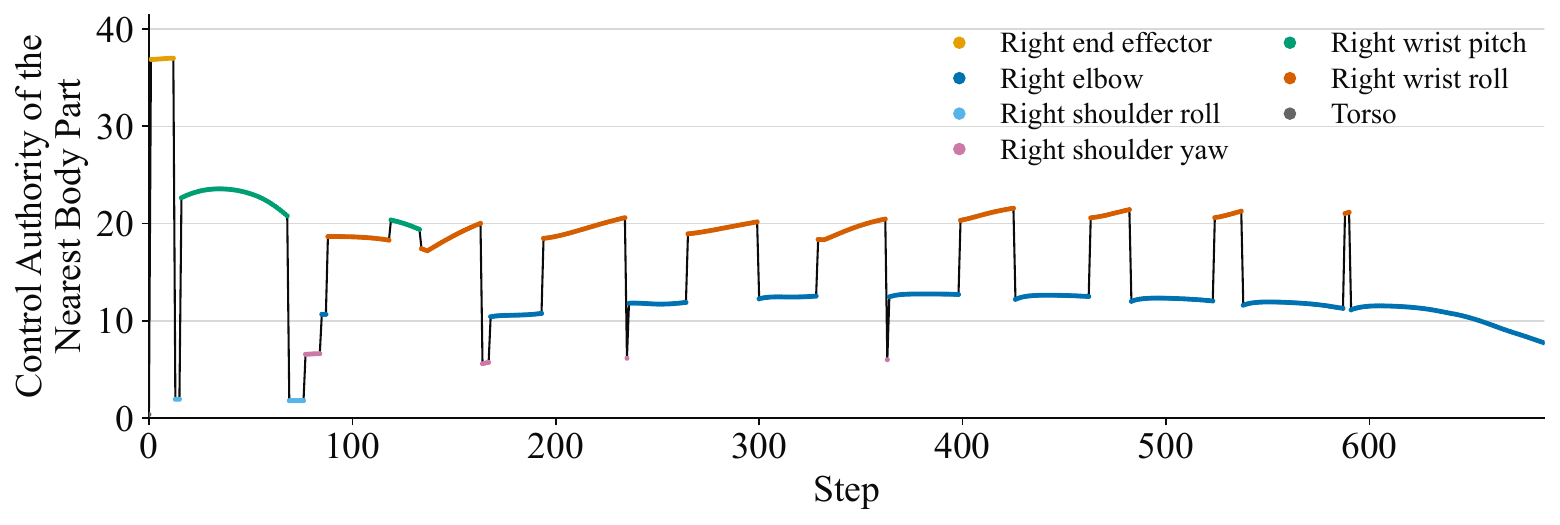}
  \caption{Control authority of the nearest body part at each step.}
\label{fig:eta20-authority}
\end{figure*}

Furthermore, we enumerate the value of $\eta$ and see how the change of the value affect the attack success on this \insertion, as shown in Figure~\ref{fig:eta-steps-task}. 
Interestingly, we observe that as we increase the value, i.e., let the robot generate controls to escape from the obstacle faster, the attack will be firstly invalidated, but become validated again after a certain value. 
This is because SSA can no longer compute a feasible solution given $\eta$'s new value.
At some states determined by the location and speed of the humanoid, the control authority of the nearest body part degrades below the expected escape rate.
Therefore, the algorithm cannot generate a control that meets the safety requirement.

Figure~\ref{fig:eta20-authority} shows this effect for $\eta_{SSA}=20$ by plotting the control authority associated with the body part closest to the obstacle at each step.
In this specific case, the nearest point to the obstacle switches across the arm, including from wrist joints to the elbow.
Because these body parts provide different control authority, the switch can leave SSA unable to compute a safe action that satisfies the configured escape rate.
It is the \insertion that induces the nearest-point change and triggers the infeasibility.

Overall, to defend against this particular attacking point, the safety escaping rate has to be carefully chosen within a niche range. 
Deciding safety escaping rate is tricky in real-world scenarios.
The range of a proper safety escaping rate is a narrow interval, and that interval differs for different attacking point, tasks, and obstacles.
Even worse, if the obstacle is dynamic or even unknown to the developer, the developer cannot reason about the proper rate range.
There may not be a proper configuration that can defend against all attacking points.
Due to the specificality and sensitivity of the parameters, the chance that an \insertion exists can be high, and many \safecontrol will be affected by such type of attack.

\subsection{Exploitation on a Real Humanoid}
\label{sec:rq2}

In this section, we deploy the identified attacks from the above experiment in the real world and validate if the identified attacks work in real-world setups.
Ideally, one would validate all identified attacks.
However, this is intractable for real-humanoid testing for the following reasons.
First, the current humanoid safety benchmark SPARK has only been properly tested for upper-body sim-to-real deployment.
Second, for the dynamic obstacle scenarios in benchmark, it will be hard to achieve accurate sim-to-real setup because it is hard to realize multiple moving obstacles each of which moves under an exact speed.
Third, such a safety experiment is risky and costly, especially that our goal is to demonstrate collision or sudden stops, both of which may cause unfixable damage to the testing humanoid and cost \$50K.

Due to the above limitations, we selected attacks that use the upper-body model and static obstacle scenarios, and we randomly chose one attacking scenario for each \safecontrol algorithm.
For each selected scenario, we replayed the attack in simulation and then reproduced the same setup on the real humanoid by matching the goal sequence and obstacle placement.

\begin{table*}[t]
\centering
\small
\setlength{\tabcolsep}{2pt}
\renewcommand{\arraystretch}{1.05}
\caption{The deployment of \insertionattack to real humanoids.}
\smallskip
\label{tab:rq2-figure-grid}
\begin{tabular}{@{}cccccccc@{}}
\toprule
\multicolumn{2}{c}{\textbf{SSA}} &
\multicolumn{2}{c}{\textbf{CBF}} &
\multicolumn{2}{c}{\textbf{SSS}} &
\multicolumn{2}{c}{} \\
\cmidrule(lr){1-2}\cmidrule(lr){3-4}\cmidrule(lr){5-6}
\textbf{Simulation} & \textbf{Real Humanoid} &
\textbf{Simulation} & \textbf{Real Humanoid} &
\textbf{Simulation} & \textbf{Real Humanoid} &
& \\
\includegraphics[width=0.115\textwidth]{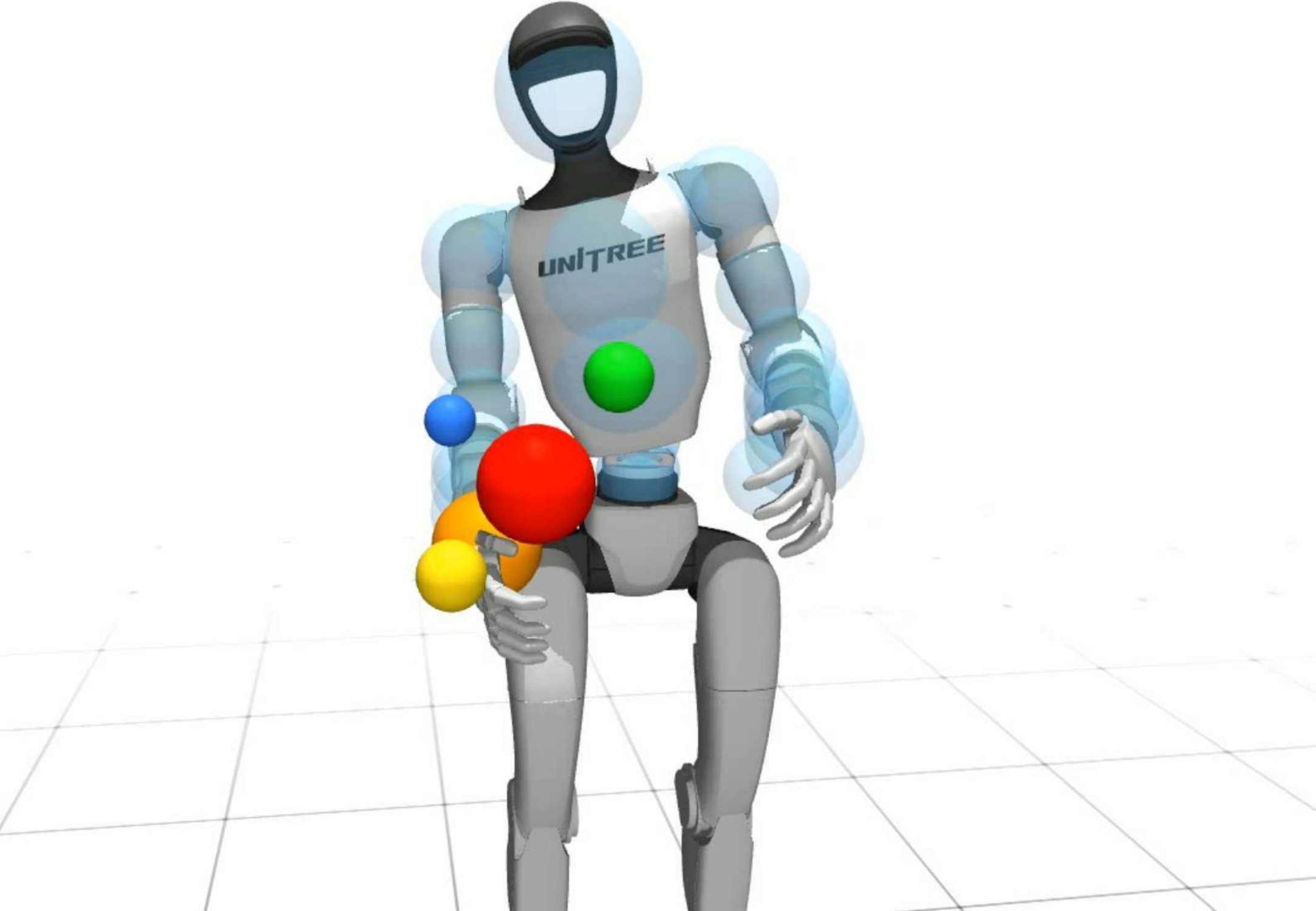} &
\includegraphics[width=0.115\textwidth]{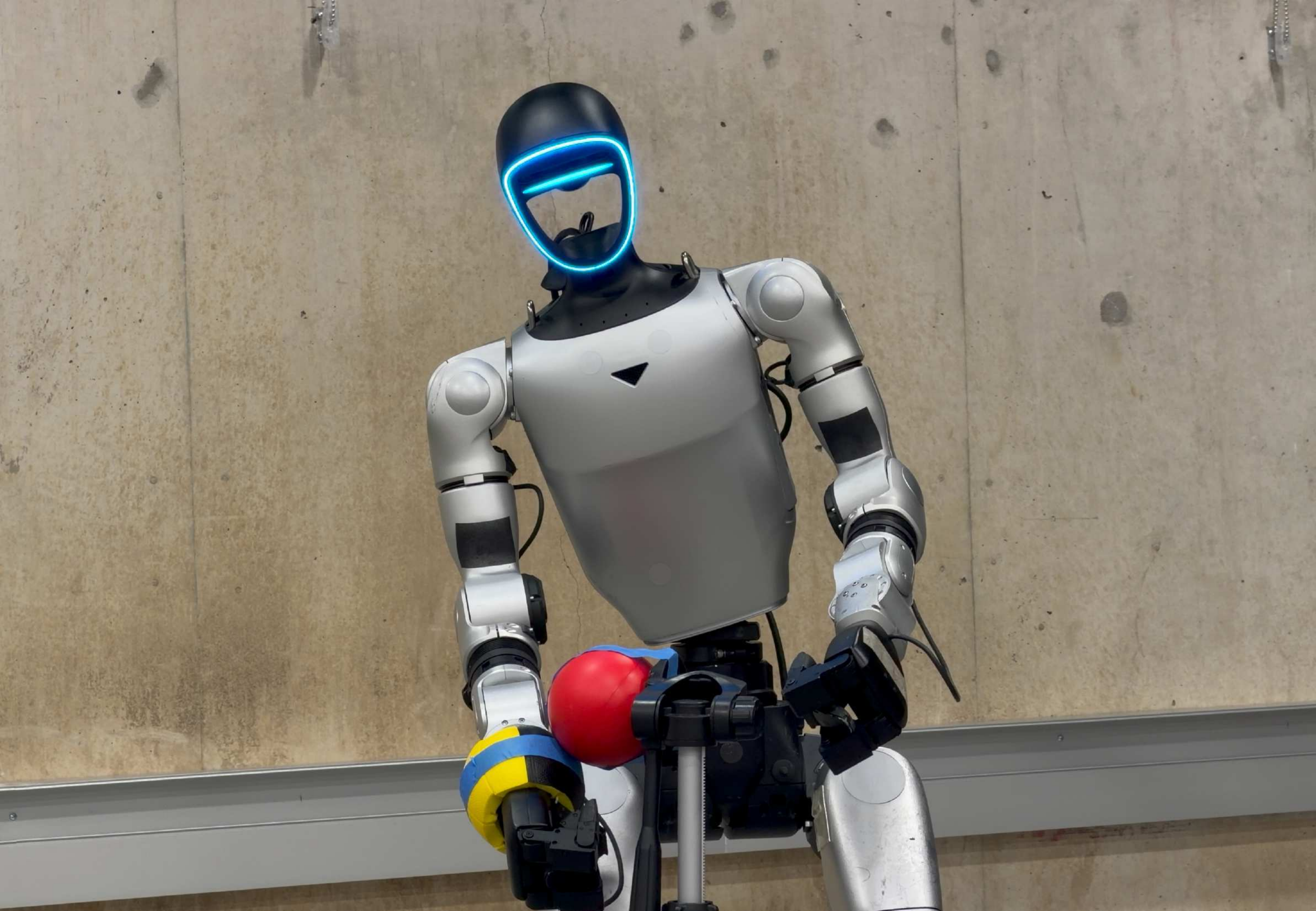} &
\includegraphics[width=0.115\textwidth]{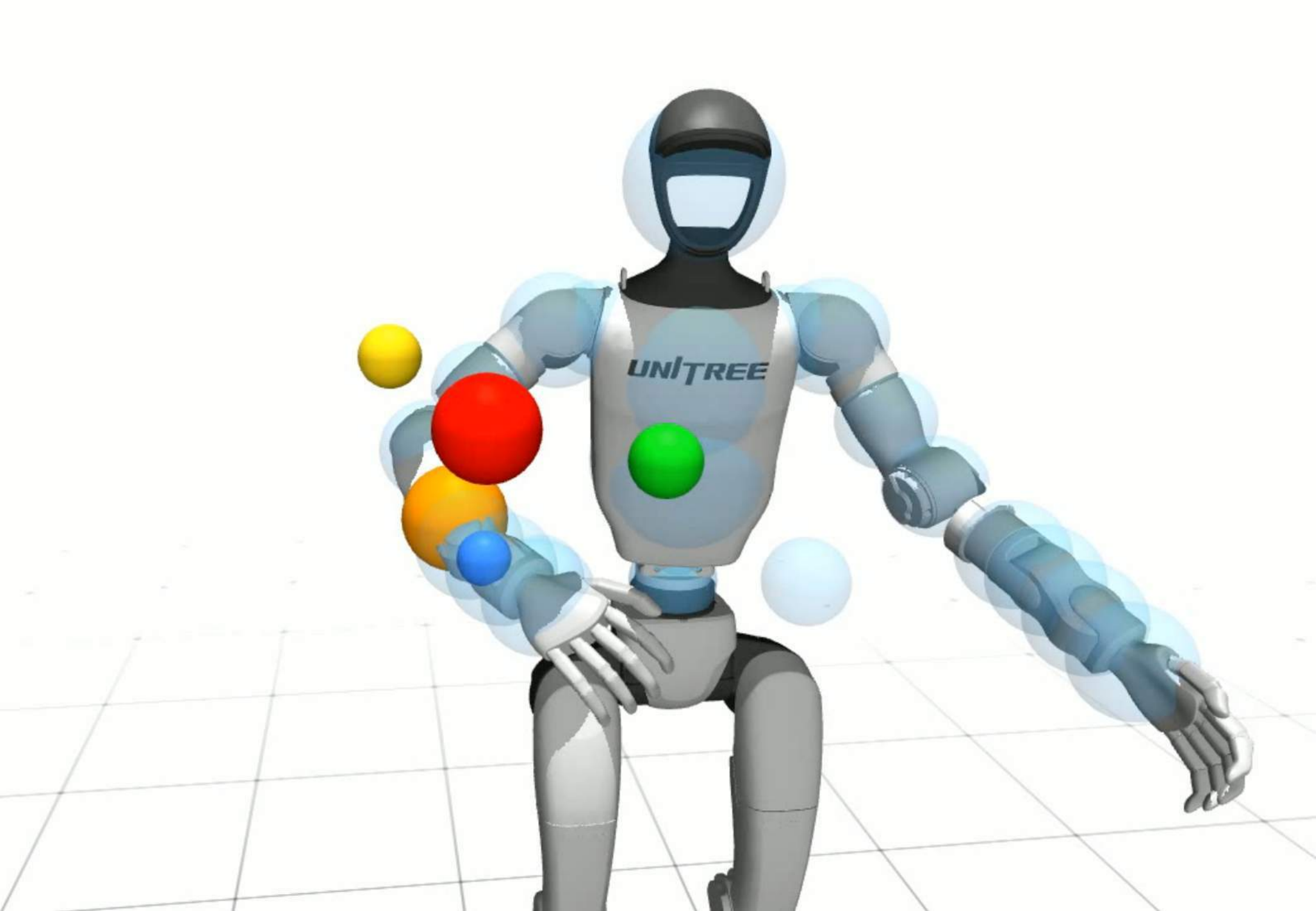} &
\includegraphics[width=0.115\textwidth]{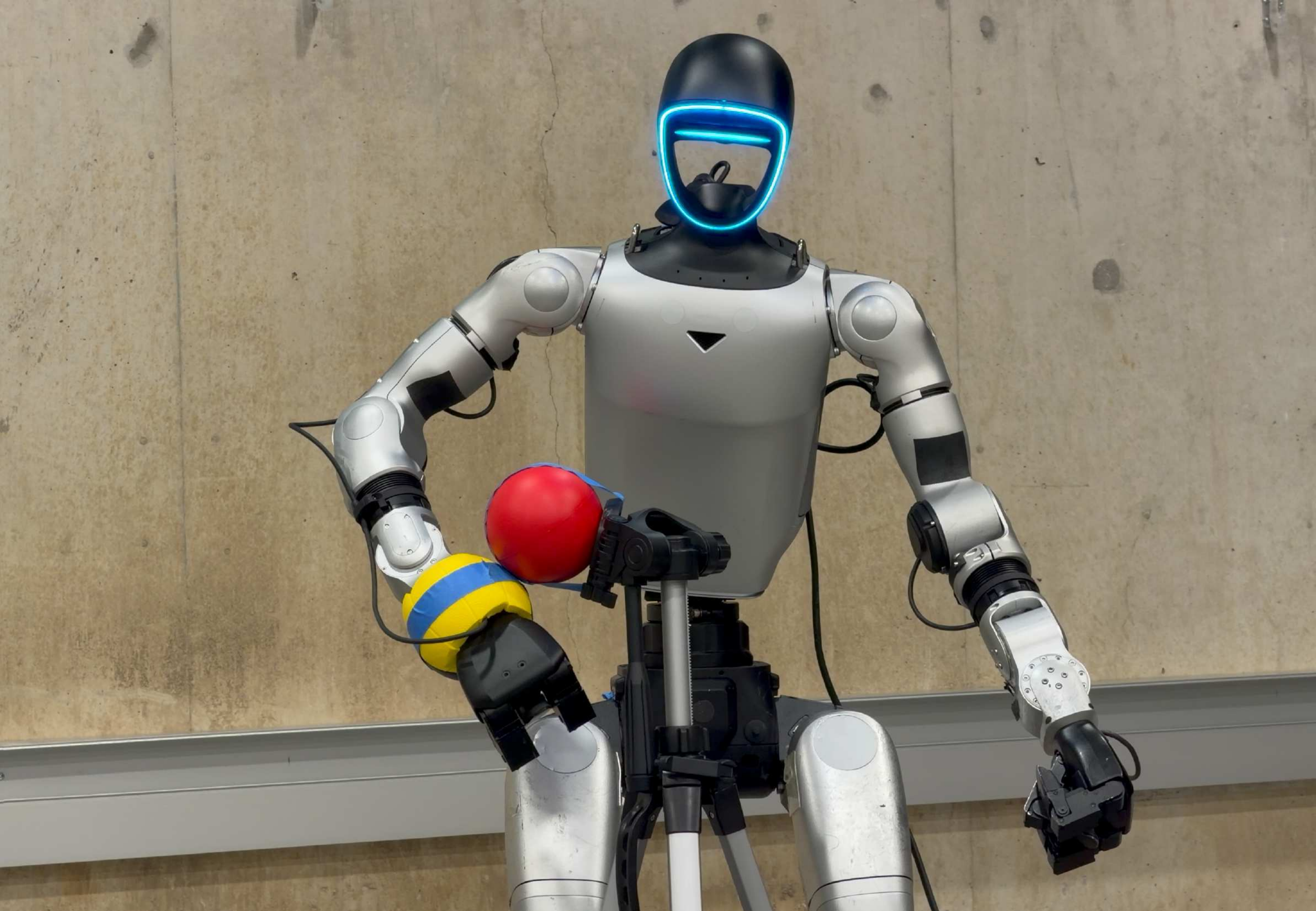} &
\includegraphics[width=0.115\textwidth]{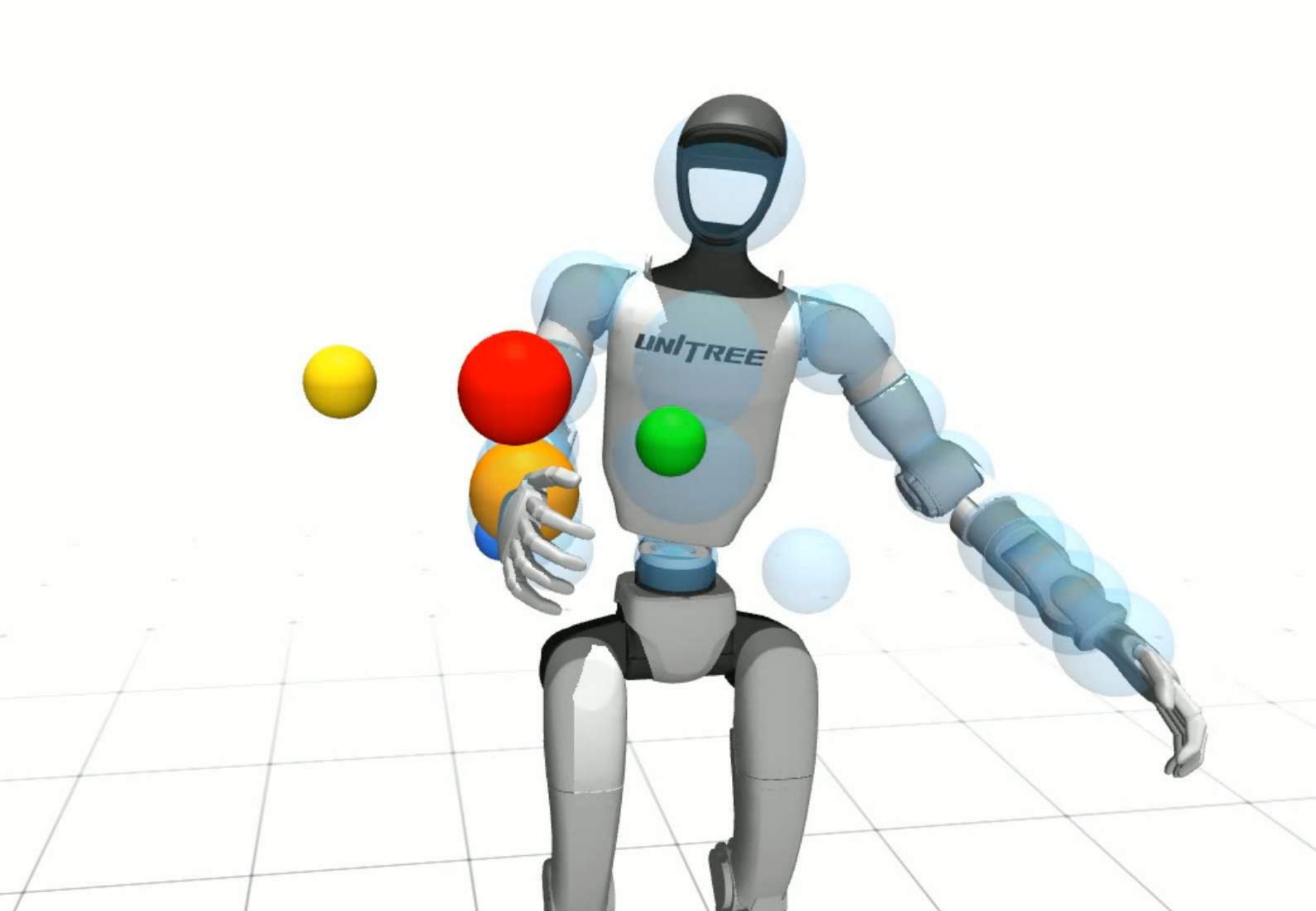} &
\includegraphics[width=0.115\textwidth]{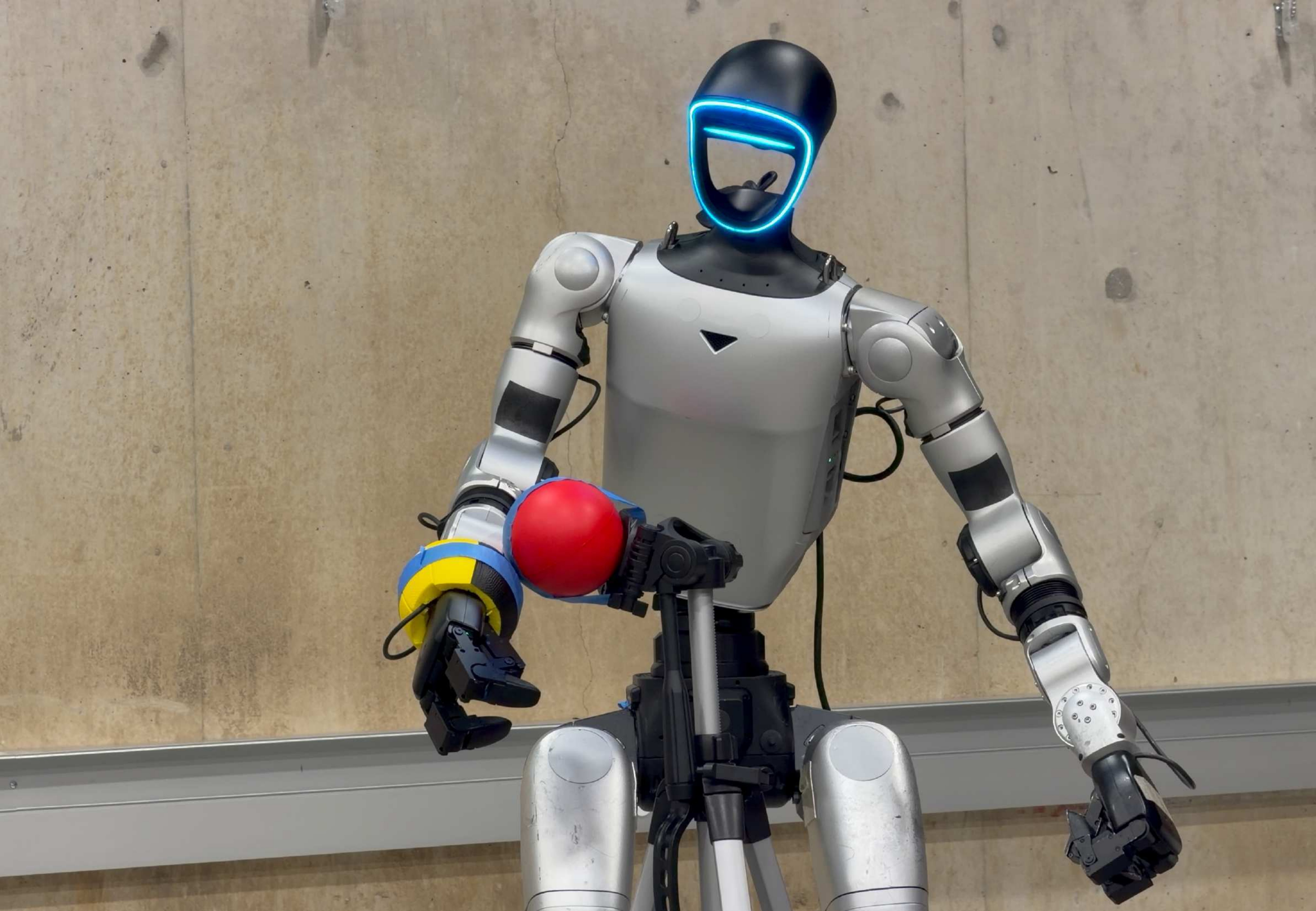} &
& \\
\midrule
\multicolumn{2}{c}{\textbf{R-SSA}} &
\multicolumn{2}{c}{\textbf{R-CBF}} &
\multicolumn{2}{c}{\textbf{R-SSS}} &
\multicolumn{2}{c}{\textbf{p-SSA}} \\
\cmidrule(lr){1-2}\cmidrule(lr){3-4}\cmidrule(lr){5-6}\cmidrule(l){7-8}
\textbf{Simulation} & \textbf{Real Humanoid} &
\textbf{Simulation} & \textbf{Real Humanoid} &
\textbf{Simulation} & \textbf{Real Humanoid} &
\textbf{Simulation} & \textbf{Real Humanoid} \\
\includegraphics[width=0.115\textwidth]{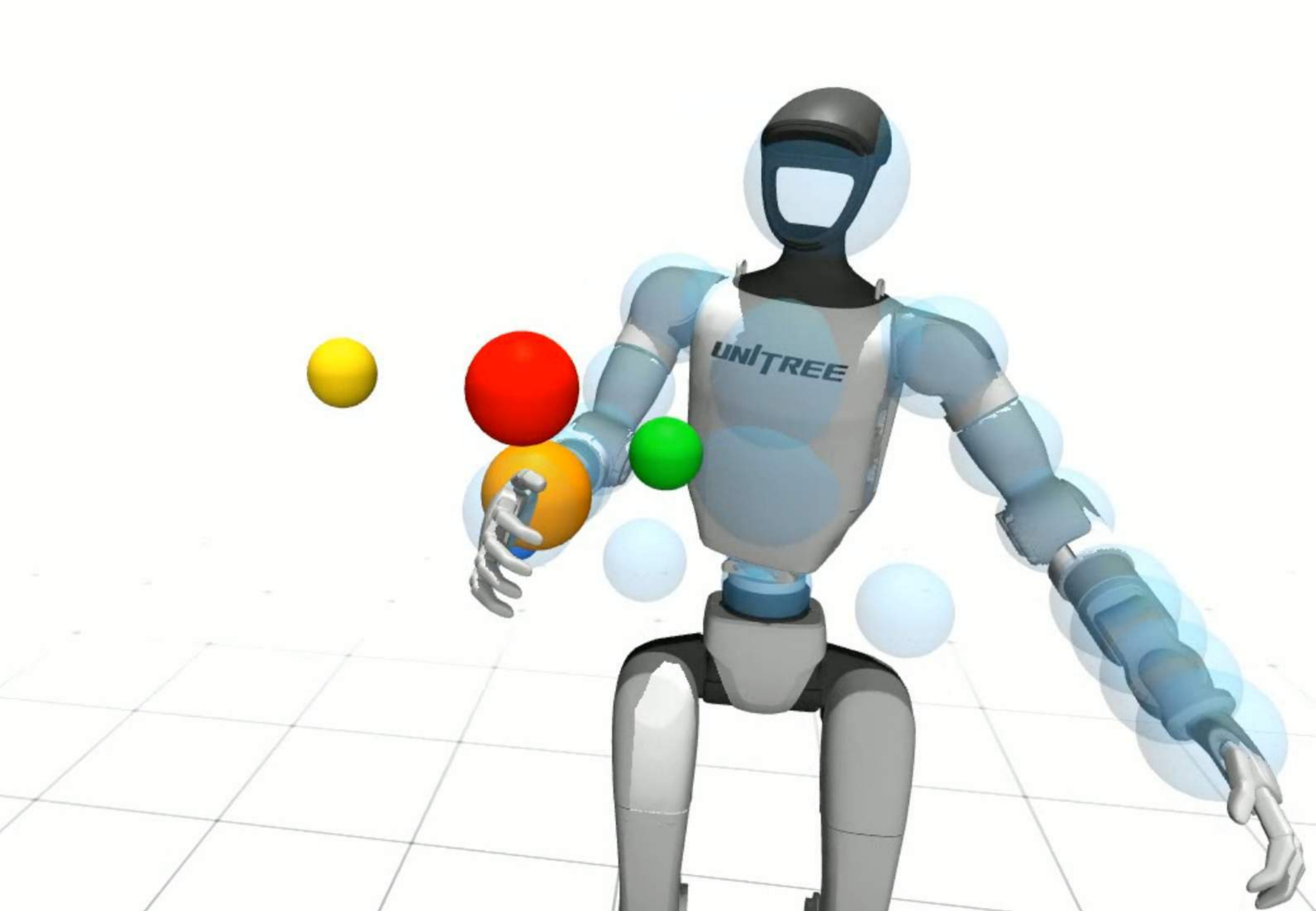} &
\includegraphics[width=0.115\textwidth]{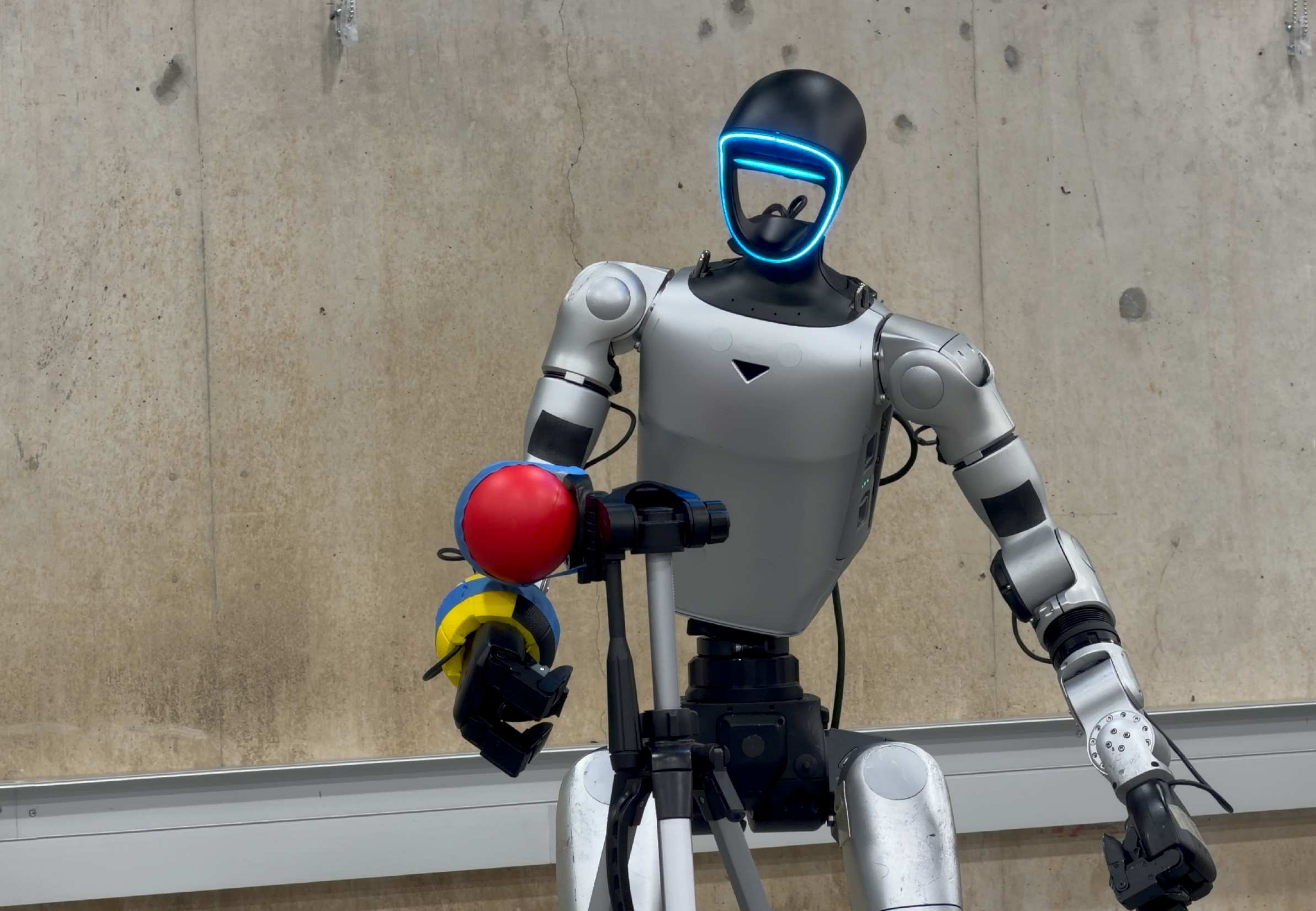} &
\includegraphics[width=0.115\textwidth]{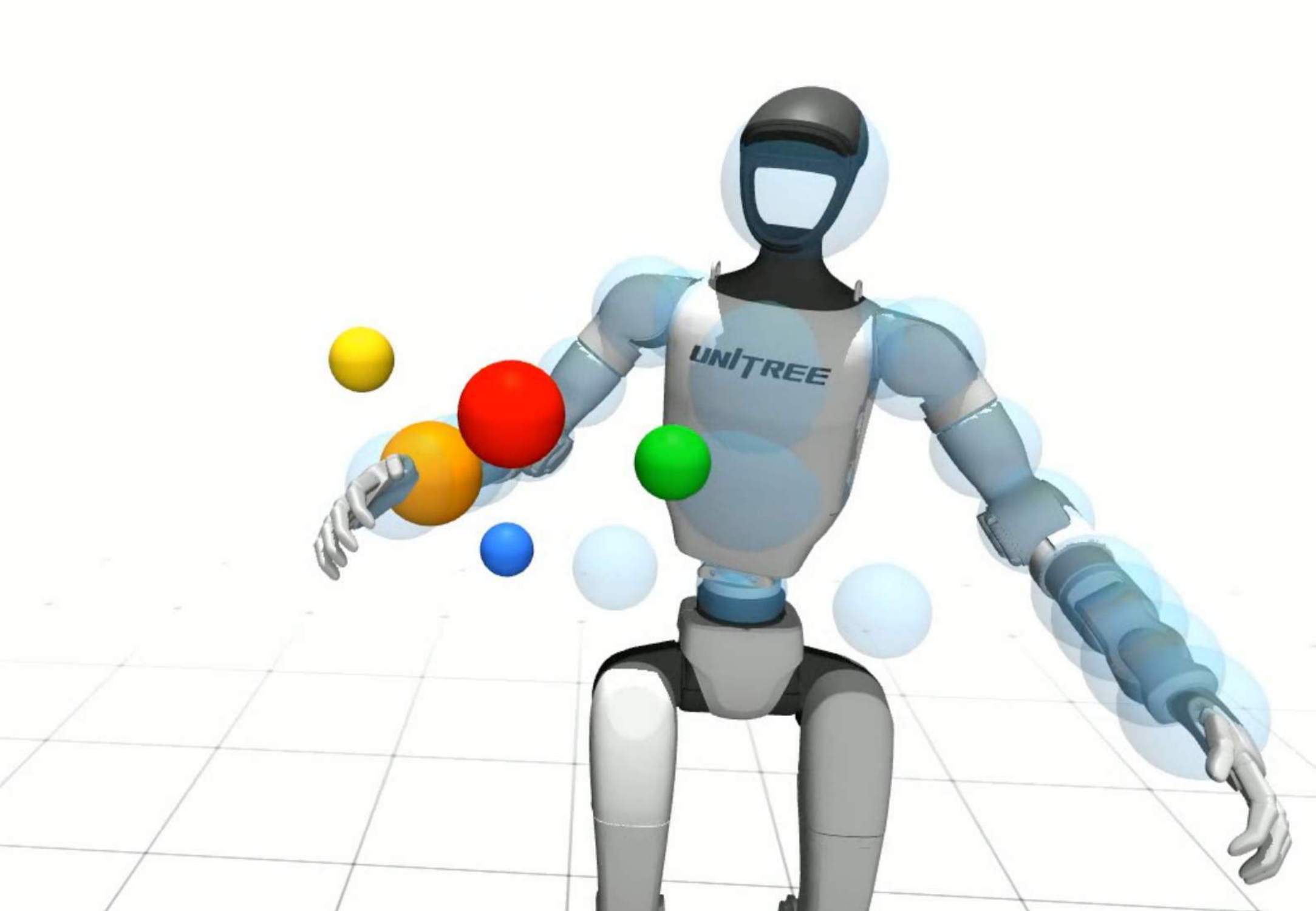} &
\includegraphics[width=0.115\textwidth]{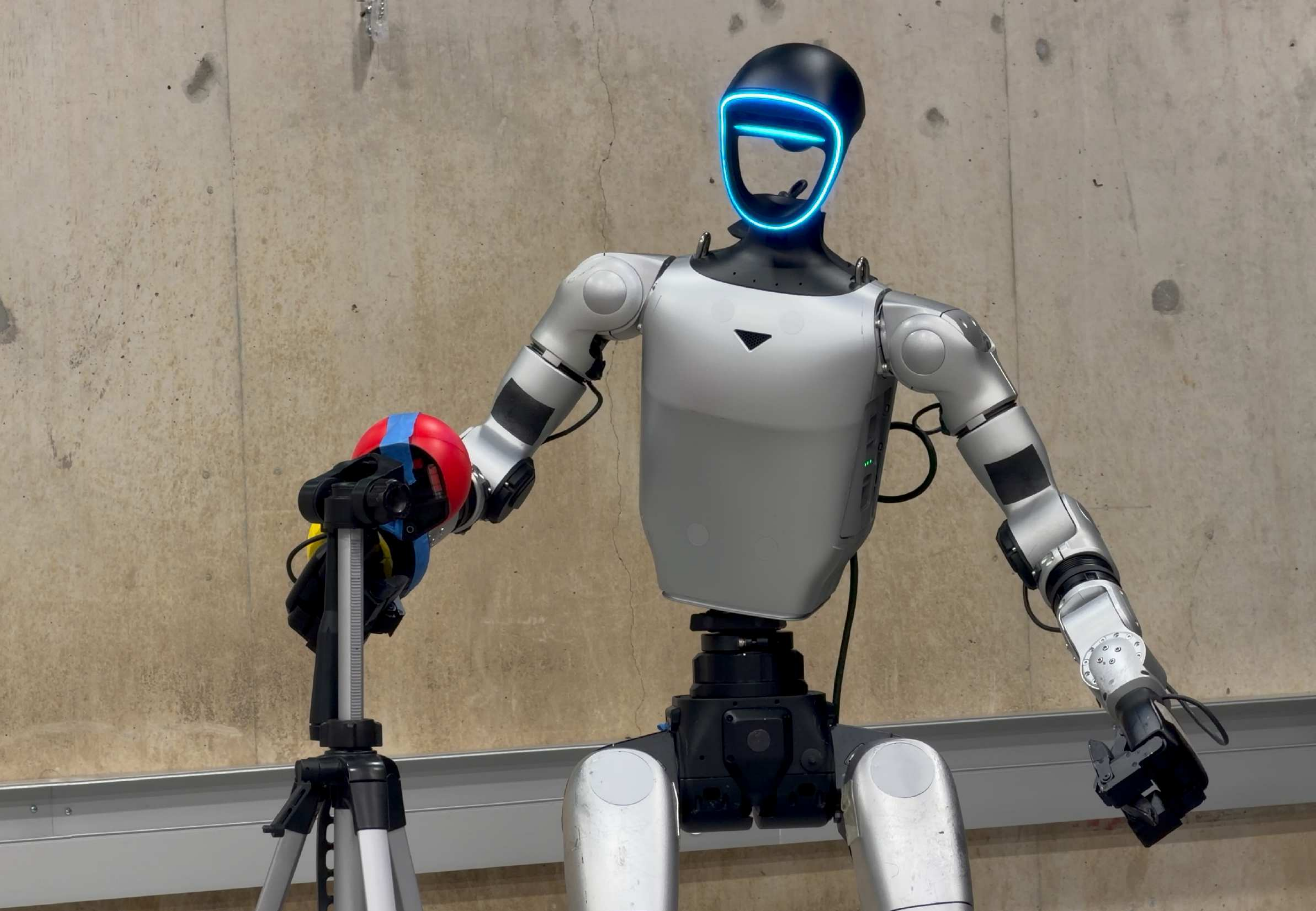} &
\includegraphics[width=0.115\textwidth]{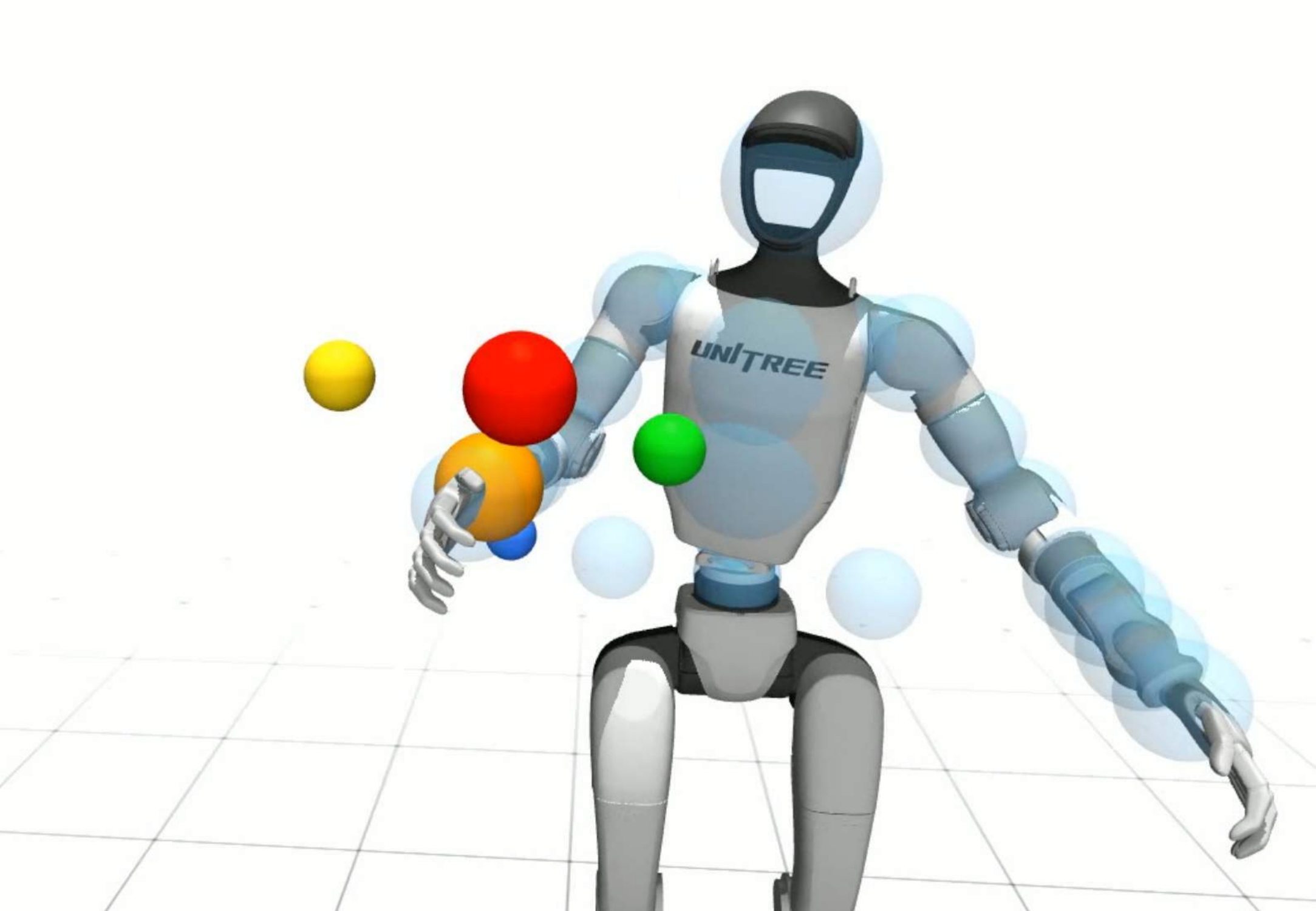} &
\includegraphics[width=0.115\textwidth]{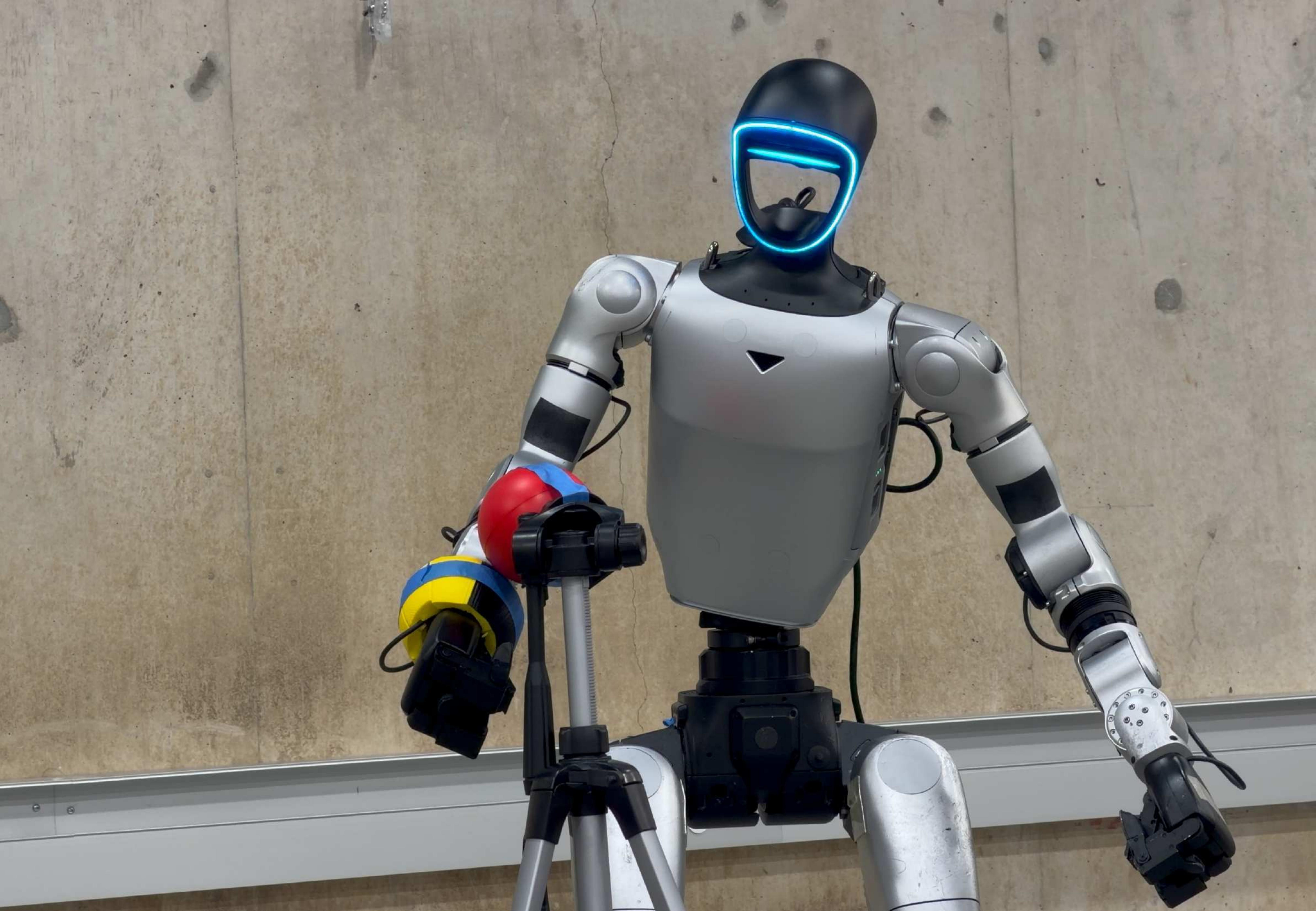} &
\includegraphics[width=0.115\textwidth]{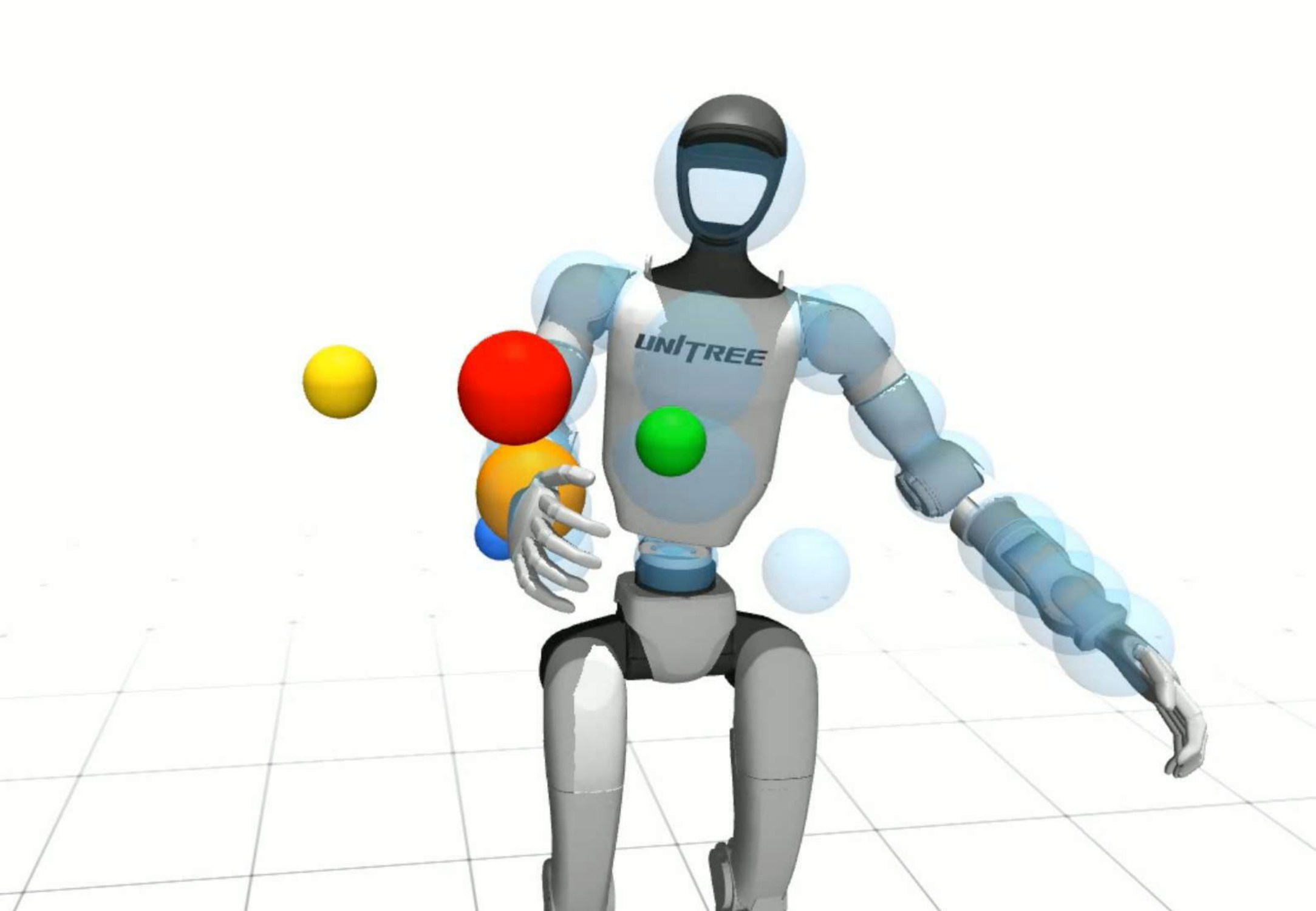} &
\includegraphics[width=0.115\textwidth]{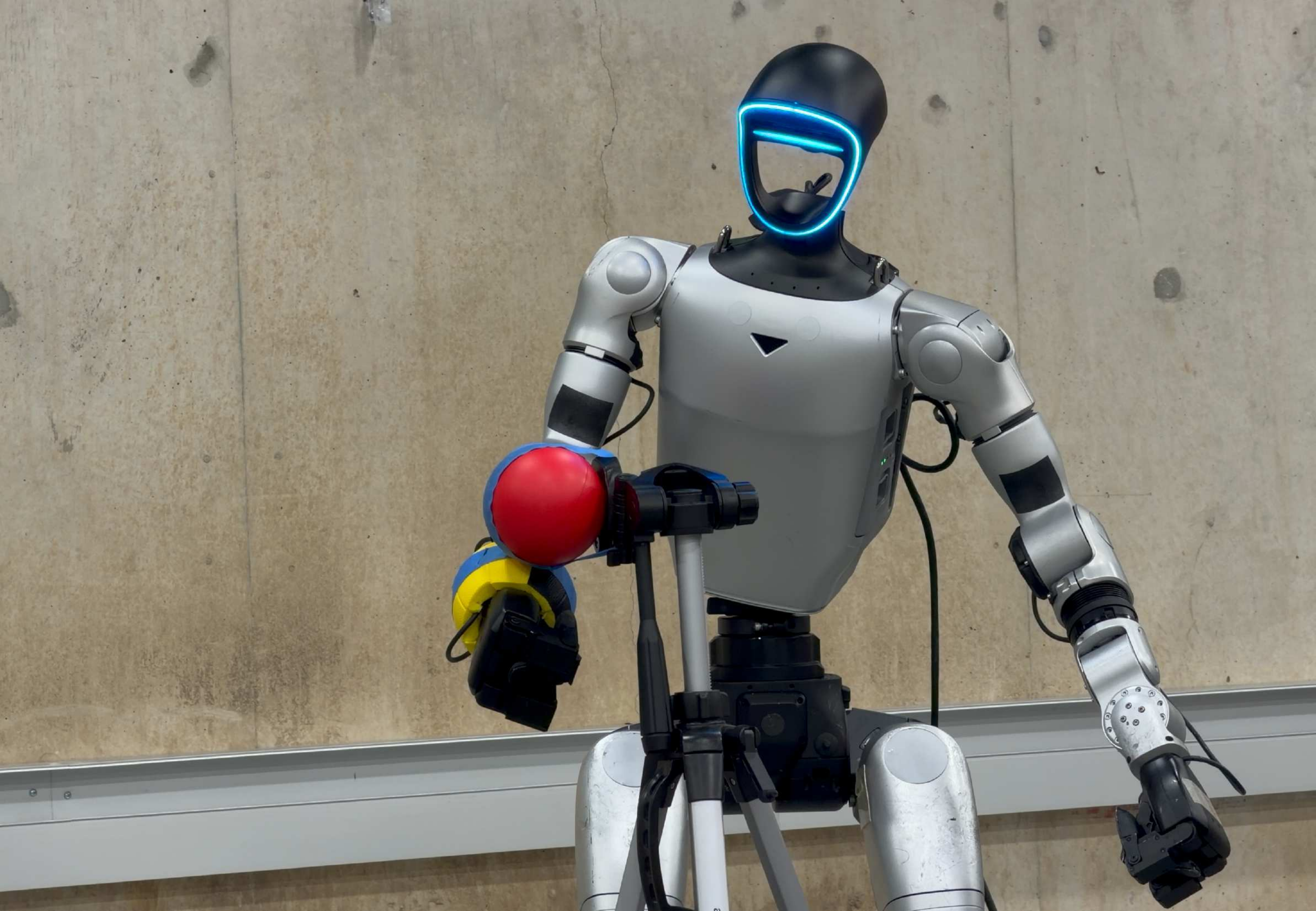} \\
\bottomrule
\end{tabular}
\end{table*}

Table~\ref{tab:rq2-figure-grid} shows the simulation and real-humanoid outcomes for all evaluated \safecontrols.
Across SSA, CBF, SSS, and their robust or relaxed variants, the same qualitative failure observed in simulation is reproduced on the physical humanoid.
This result indicates that the attacks identified in the controlled simulation setting are not merely simulator artifacts: when the environment is restricted to the upper-body, single-obstacle setting that can be faithfully reproduced on hardware, the attack transfers to the real robot.

\begin{figure*}[t]
\centering
\includegraphics[width=0.9\textwidth]{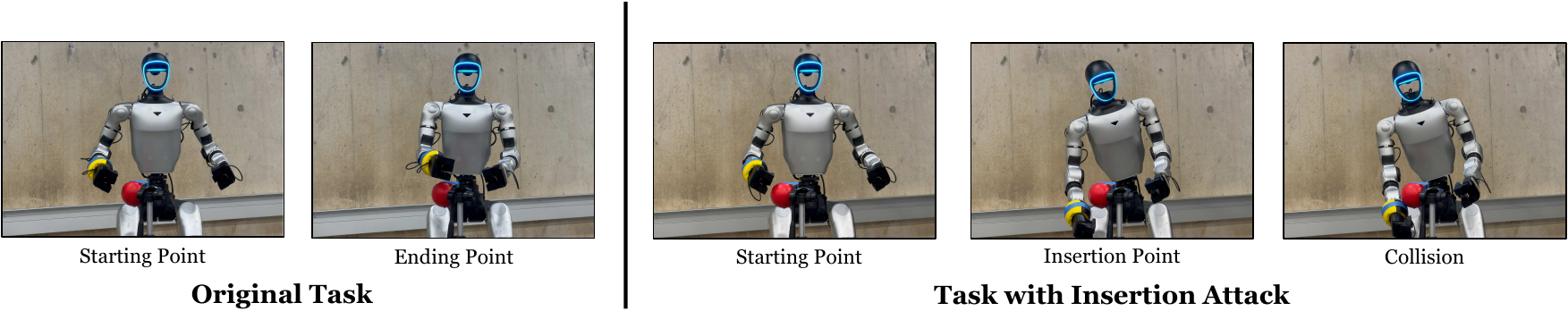}
\caption{Real-world attack demonstration. The original task succeeds without contacting the obstacle, while the insertion attack drives the humanoid through an intermediate point and causes a collision during the return to the original goal.}
\label{fig:rq2-ssa-comparison}
\end{figure*}

To further explain the sim-to-real deployment procedure, we present one representative case study.
Using the SSA scenario as an example, Figure~\ref{fig:rq2-ssa-comparison} compares the original task with the attacked execution on the deployed real-world setup.
In the original task, the humanoid successfully moves from the starting point to the ending point without contacting the obstacle.
With the \insertion attack, however, the inserted goal first drives the arm toward the obstacle; when the humanoid subsequently moves toward the original ending point, the arm collides with the obstacle.
This case study illustrates the same validation condition used throughout RQ2: the task is executable in the real setup without the insertion, but the same setup fails once the attacker inserts the intermediate goal.

\subsection{Attack Discovery Effectiveness}
\label{sec:rq3}

In this experiment, we evaluate the effectiveness of the CEM-based search in identifying \insertions in vulnerable scenarios.
Unlike the previous experiment, which establishes the existence of \theattack, this experiment evaluates its scalability.
Specifically, we measure how many successful attacks an automated search algorithm can discover in a vulnerable scenario.

To this end, we construct a dataset of vulnerable scenarios from the SPARK benchmark.
We randomly sample scenarios from the benchmark and validate each scenario against a \safecontrol.
A scenario is considered vulnerable if the original task sequence $[\mathbf{G}_0,\mathbf{G}_1]$ succeeds, while a planted sequence $[\mathbf{G}_0,\mathbf{G}',\mathbf{G}_1]$ consistently causes a failure during the second task.
This process yields five vulnerable scenarios for each \safecontrol, for a total of 35 scenarios.

For each scenario, we compare the following search strategies:
\begin{itemize}[nosep]
\item \emph{Random}: uniformly samples insertion points in the three-dimensional search space.
\item \emph{\tool}: the proposed CEM-based search algorithm.
\end{itemize}

We also evaluate two variants, \emph{Random-NI} and \emph{\tool-NI}, which initialize the search near obstacles. These variants test the hypothesis that successful \insertions are more likely to occur close to obstacles.

Simulation time varies across runs, and simulation dominates the overall search cost.
Therefore, instead of limiting wall-clock time, we assign every search algorithm the same budget of 125 simulation trials.
This design provides a fair comparison across search strategies.
On average, each search process costs \averagetime minutes.

\begin{table}[t]
  \centering
  \small
  \caption{Search effectiveness across \datasize scenarios.}
  \smallskip
  \label{tab:rq3}
  \begin{tabular}{lrrrr}
  \hline
  \textbf{Method} &
  \textbf{\begin{tabular}[c]{@{}c@{}}\# Attacked\\Scenarios\end{tabular}} &
  \textbf{\begin{tabular}[c]{@{}c@{}}Scenario\\Hit Rate\end{tabular}} &
  \textbf{\begin{tabular}[c]{@{}c@{}}\# Total\\Attacks\end{tabular}} &
  \textbf{\begin{tabular}[c]{@{}c@{}}Attack\\Hit Rate\end{tabular}} \\
  \hline
  Random    & 40/40 & 100.0\% & 1760 & 35.2\% \\
  Random-NI & 40/40 & 100.0\% & 1751 & 35.0\% \\
  SIREN     & 40/40 & 100.0\% & 3277 & 65.5\% \\
  SIREN-NI  & 38/40 & 95.0\%  & 3006 & 60.1\% \\
  \hline
  \end{tabular}
\end{table}

We synthesize the experiment results shown in Table~\ref{tab:rq3}.
Both random search and \tool identify at least one successful attack in all \datasize scenarios.
However, \tool discovers substantially more attacks than random search: \tool identifies 3277 attacks, whereas the random search approach identifies 1760 attacks, which is nearly only a half of \tool's discovery.
The attack hit rates of \tool and \tool-NI both exceed 60\%.
Surprisingly, neither near-obstacle initialization variant outperforms its corresponding baseline.
Although \tool-NI achieves a slightly higher hit rate, it fails to identify attacks in every scenario.
These results demonstrate that \tool is significantly more effective than random search for identifying attacks, while providing no evidence that successful \insertions are concentrated near obstacles.

\begin{figure}[t]
\centering
\includegraphics[width=0.6\columnwidth]{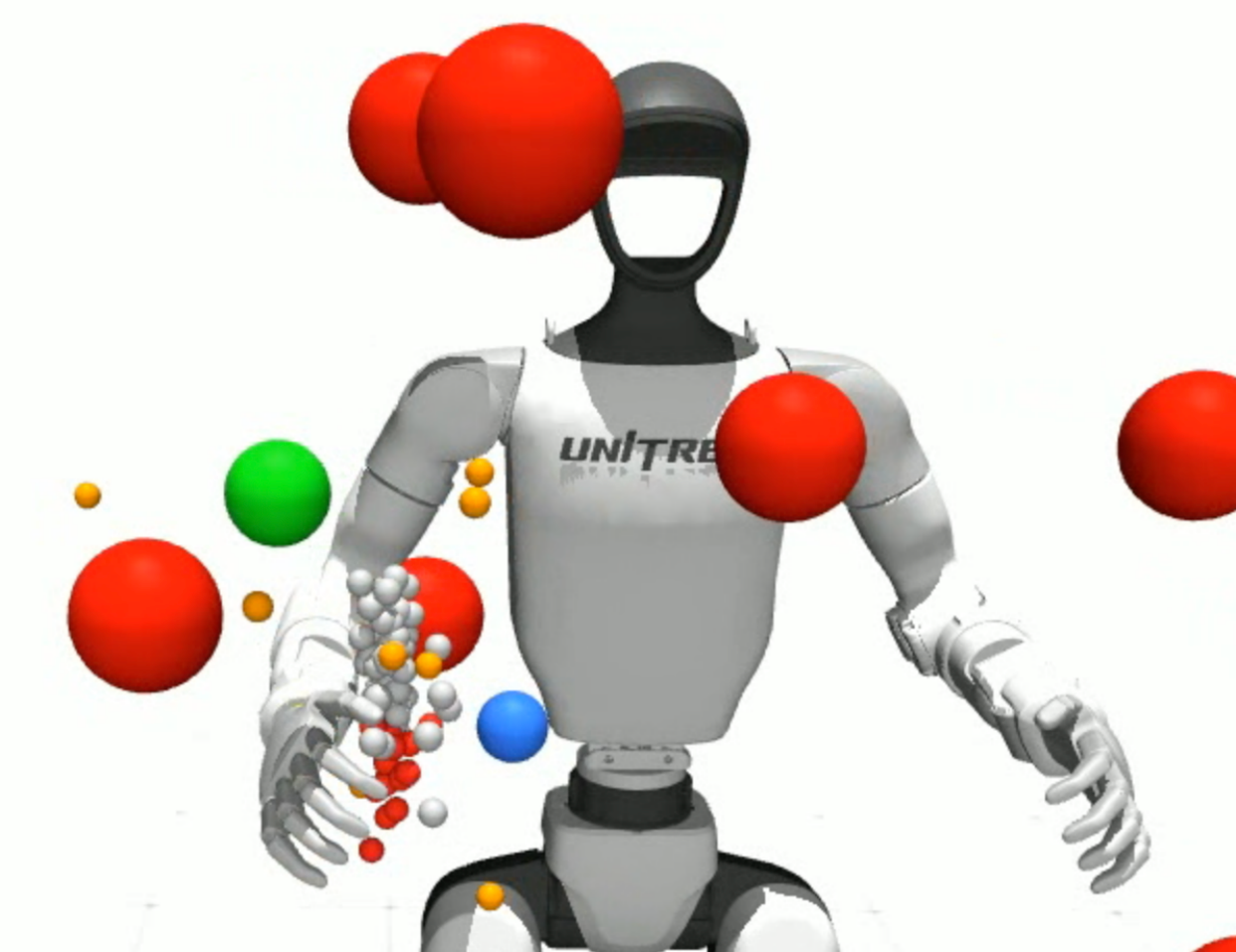}
\caption{Case study: searched points by \tool. Large red balls denote the obstacles. The blue point is the starting point, and the green point is the ending point of the original task.}
\label{fig:siren}
\end{figure}

Specifically, we examine the search process of \tool, as illustrated in Figure~\ref{fig:siren}.
In the figure, the small red points denote the \insertions identified by \tool, whereas the white points represent all sampled search points.
\tool initializes the search with points sampled uniformly at random in the three-dimensional workspace.
During the CEM-based optimization, the search rapidly converges to the region between the two obstacles, where it identifies 21 attacking points out of 125 search trials.
This result shows that CEM efficiently concentrates the search on regions that are likely to contain attacking points.
We also observe that points located near obstacles are not necessarily attacking points.
This observation explains why initializing the search near obstacles does not improve attack discovery.

\section{Potential Defense and Mitigation}

The root cause of the vulnerability in \safecontrol is its design.
Safety parameters depend on both the task and the environment, and inaccurate parameter configurations create opportunities for attack.
Unfortunately, no existing method can determine the correct parameter values for arbitrary tasks and environments.
Moreover, little prior work studies the relationship between safety parameters and task or environmental characteristics.

A complete defense against \theattack requires accurate parameter configurations for each task and environment while keeping both fixed during execution.
An alternative direction is to develop new safety algorithms that adapt their parameters dynamically according to changes in the task or environment~\cite{yun2025siadaptation,dawson2023safecert,zhao2021modelfree}.

One practical mitigation is to prohibit insertion into tasks that have already been scheduled.
This restriction prevents failures of scheduled tasks.
Because the attacker cannot predict future tasks, it also makes it significantly more difficult to insert goals at the precise time required for a successful attack.
Although this mitigation may reduce the flexibility and efficiency of the humanoid, it improves both task reliability and operational safety.

Another possible mitigation is to introduce an additional runtime monitor that halts the humanoid upon collision.
However, such a mechanism requires more than a conventional emergency stop.
Existing emergency-stop procedures or transitions to a fixed recovery pose can themselves cause the humanoid to fall or collide during recovery~\cite{emergencystop}.
Therefore, defenders need a dynamic emergency-stop mechanism that minimizes collateral damage while safely terminating execution.

Although mitigations reduce the attack surface, they cannot eliminate the vulnerability fundamentally.
Therefore, this limitation is structural; as long as \safecontrols compute controls solely from the current state without reasoning about future feasibility or permitting temporary task-reversing maneuvers, an adversary can always steer the robot toward a state that causes safety infeasibility.
Eliminating this vulnerability requires predictive safe control with inclusive feasibility guarantees, which remains an open research problem~\cite{zhao2023sis,wabersich2021psf,gurriet2018activeset,mitchell2005hj}.

\section{Limitations and Future Work}

Although we show that \theattack is possible, we do not know the prevalence of \theattack.
In particular, given an arbitrary task and environment, we cannot estimate the probability that a successful attack exists.
Quantifying this prevalence is challenging because the state space of humanoid systems is prohibitively large.
Although many scenarios do not admit a successful attack, we currently lack a principled method to determine whether an attack exists or to characterize the conditions under which one becomes feasible.

Also, this paper assumes that the attacker has complete knowledge of the safety filter deployed on the target humanoid, including its algorithm and parameter values.
Future work could relax this assumption by considering attackers with incomplete information.
For example, an attacker may know the type of \safecontrol but not its parameter values, or may not know the deployed \safecontrol at all.
With limited interaction or probing of the target robot, it may be possible to infer sufficient information to construct effective attacks under these weaker assumptions.

Lastly, this paper does not consider the complete humanoid software stack or its impact on attack success.
A practical humanoid system includes components such as perception and simultaneous localization and mapping (SLAM), which estimate the robot's pose and the locations of surrounding obstacles.
Errors in these components may affect the execution and success of \theattack.
As a first step, this paper focuses on demonstrating that the vulnerability in \safecontrol is fundamentally exploitable.
Future work will investigate attacks against end-to-end humanoid systems that remain effective despite inaccuracies in perception, localization, and other system components.

\section{Conclusion}

In this paper, we demonstrate that existing \safecontrol algorithms contain security vulnerabilities that attackers can exploit to cause safety violations such as collisions.
We present \theattack and show that a scalable and automated approach can identify vulnerable \insertions, which underscores the practical severity of the vulnerability.
Our evaluation confirms that \theattack exists in realistic scenarios and can compromise real humanoid robots.
Finally, we discuss potential defenses and mitigation strategies against \theattack.

\clearpage
\bibliographystyle{plainurl}
\bibliography{references}

@inproceedings{ssa,
  author    = {Liu, Changliu and Tomizuka, Masayoshi},
  title     = {Algorithmic Safety Measures for Intelligent Industrial Co-Robots},
  booktitle = {IEEE International Conference on Robotics and Automation (ICRA)},
  pages     = {3095--3102},
  year      = {2016},
  doi       = {10.1109/ICRA.2016.7487476}
}

@article{cbf,
  author        = {Ames, Aaron D. and Xu, Xiangru and Grizzle, Jessy W. and Tabuada, Paulo},
  title         = {Control Barrier Function Based Quadratic Programs for Safety Critical Systems},
  journal       = {IEEE Transactions on Automatic Control},
  volume        = {62},
  number        = {8},
  pages         = {3861--3876},
  year          = {2017},
  doi           = {10.1109/TAC.2016.2638961},
  eprint        = {1609.06408},
  archivePrefix = {arXiv}
}

@inproceedings{wei2019energy,
  author    = {Wei, Tianhao and Liu, Changliu},
  title     = {Safe Control Algorithms Using Energy Functions: A Unified Framework,
               Benchmark, and New Directions},
  booktitle = {IEEE Conference on Decision and Control (CDC)},
  year      = {2019},
  note      = {arXiv:1908.01883}
}

@article{pssa,
  author        = {Chen, Rui and Sun, Yifan and Liu, Changliu},
  title         = {Dexterous Safe Control for Humanoids in Cluttered Environments
                   via Projected Safe Set Algorithm},
  journal       = {arXiv preprint arXiv:2502.02858},
  year          = {2025},
  eprint        = {2502.02858},
  archivePrefix = {arXiv},
  primaryClass  = {cs.RO},
  url           = {https://arxiv.org/abs/2502.02858}
}

@inproceedings{spark,
  author    = {Sun, Yifan and Chen, Rui and Yun, Kai S. and Fang, Yikuan and
               Jung, Sebin and Li, Feihan and Li, Bowei and Zhao, Weiye and Liu, Changliu},
  title     = {{SPARK}: Safe Protective and Assistive Robot Kit},
  booktitle = {IFAC Symposium on Robotics},
  year      = {2025},
  eprint    = {2502.03132},
  archivePrefix = {arXiv},
  primaryClass  = {cs.RO},
  url       = {https://arxiv.org/abs/2502.03132}
}

@inproceedings{zhao2023sis,
  author        = {Zhao, Weiye and He, Tairan and Wei, Tianhao and Liu, Simin and Liu, Changliu},
  title         = {Safety Index Synthesis via Sum-of-Squares Programming},
  booktitle     = {American Control Conference (ACC)},
  pages         = {732--737},
  year          = {2023},
  organization  = {IEEE},
  doi           = {10.23919/ACC55779.2023.10156463},
  eprint        = {2209.09134},
  archivePrefix = {arXiv},
  primaryClass  = {eess.SY}
}

@inproceedings{agrawal2021iccbf,
  author    = {Agrawal, Devansh K. and Panagou, Dimitra},
  title     = {Safe Control Synthesis via Input Constrained Control Barrier Functions},
  booktitle = {IEEE Conference on Decision and Control (CDC)},
  year      = {2021},
  note      = {arXiv:2104.01704}
}

@inproceedings{chen2021backupcbf,
  author    = {Chen, Yuxiao and Jankovic, Mrdjan and Santillo, Mario and Ames, Aaron D.},
  title     = {Backup Control Barrier Functions: Formulation and Comparative Study},
  booktitle = {IEEE Conference on Decision and Control (CDC)},
  year      = {2021},
  note      = {arXiv:2104.11332}
}

@article{wabersich2021psf,
  author  = {Wabersich, Kim P. and Zeilinger, Melanie N.},
  title   = {A Predictive Safety Filter for Learning-Based Control of Constrained
             Nonlinear Dynamical Systems},
  journal = {Automatica},
  volume  = {129},
  pages   = {109597},
  year    = {2021}
}

@article{emergencystop,
  author        = {Ding, Caiwu and Cui, Tao and Wang, Lingyun and Wen, Chengtao},
  title         = {Toward Certified Functional Safety for Industrial Humanoid Robots:
                   The Fail-Passive Gap and a Feasibility Study},
  journal       = {arXiv preprint arXiv:2608.02809},
  year          = {2026},
  eprint        = {2608.02809},
  archivePrefix = {arXiv},
  primaryClass  = {cs.RO},
  url           = {https://arxiv.org/abs/2608.02809}
}

@inproceedings{jailbreak,
  author    = {Prieto-L{\'{o}}pez, L. and S{\'{a}}nchez de la Fuente, S. and
               Carrera-Mart{\'{i}}nez, D. and Rodr{\'{i}}guez-Lera, F. J. and
               S{\'{a}}nchez-Gonz{\'{a}}lez, L.},
  title     = {Cybersecurity Audit of the {Unitree Go2} Quadruped Robot},
  booktitle = {Artificial Intelligence in Security},
  publisher = {Springer},
  year      = {2026},
  doi       = {10.1007/978-3-032-29251-3_12}
}

@misc{unitreeonline,
  author       = {{Unitree Robotics}},
  title        = {Unitree Robot {SDK} Version 2 (\texttt{unitree\_sdk2})},
  howpublished = {\url{https://github.com/unitreerobotics/unitree_sdk2}},
  year         = {2024},
  note         = {Accessed 2025}
}

@book{astrom1995pid,
  author    = {{\AA}str{\"o}m, Karl Johan and H{\"a}gglund, Tore},
  title     = {{PID} Controllers: Theory, Design, and Tuning},
  edition   = {2nd},
  publisher = {Instrument Society of America},
  year      = {1995}
}

@book{siciliano2009robotics,
  author    = {Siciliano, Bruno and Sciavicco, Lorenzo and Villani, Luigi
               and Oriolo, Giuseppe},
  title     = {Robotics: Modelling, Planning and Control},
  publisher = {Springer},
  year      = {2009},
  doi       = {10.1007/978-1-84628-642-1}
}

@article{levine2016e2e,
  author  = {Levine, Sergey and Finn, Chelsea and Darrell, Trevor and Abbeel, Pieter},
  title   = {End-to-End Training of Deep Visuomotor Policies},
  journal = {Journal of Machine Learning Research},
  volume  = {17},
  number  = {39},
  pages   = {1--40},
  year    = {2016}
}

@inproceedings{brohan2023rt2,
  author        = {Brohan, Anthony and Brown, Noah and Carbajal, Justice and
                   Chebotar, Yevgen and Chen, Xi and Choromanski, Krzysztof and
                   Ding, Tianli and Driess, Danny and Dubey, Avinava and
                   Finn, Chelsea and others},
  title         = {{RT-2}: Vision-Language-Action Models Transfer Web Knowledge
                   to Robotic Control},
  booktitle     = {Conference on Robot Learning (CoRL)},
  year          = {2023},
  eprint        = {2307.15818},
  archivePrefix = {arXiv}
}

@inproceedings{kim2024openvla,
  author        = {Kim, Moo Jin and Pertsch, Karl and Karamcheti, Siddharth and
                   Mees, Oier and Bahl, Shikhar and Loquercio, Antonio and
                   Shi, Suraj and Doshi-Velez, Finale and Finn, Chelsea and
                   Levine, Sergey and others},
  title         = {{OpenVLA}: An Open-Source Vision-Language-Action Model},
  booktitle     = {Conference on Robot Learning (CoRL)},
  year          = {2024},
  eprint        = {2406.09246},
  archivePrefix = {arXiv}
}

@article{black2024pi0,
  author        = {Black, Kevin and Brown, Noah and Driess, Danny and
                   Esmail, Adnan and Equi, Michael and Finn, Chelsea and
                   Fusai, Niccolo and Groom, Lachy and Hausman, Karol and
                   Ichter, Brian and others},
  title         = {$\pi_0$: A Vision-Language-Action Flow Model for
                   General Robot Control},
  journal       = {arXiv preprint arXiv:2410.24164},
  year          = {2024},
  eprint        = {2410.24164},
  archivePrefix = {arXiv}
}

@article{rubinstein1999cem,
  author  = {Rubinstein, Reuven Y.},
  title   = {The Cross-Entropy Method for Combinatorial and Continuous Optimization},
  journal = {Methodology and Computing in Applied Probability},
  volume  = {1},
  number  = {2},
  pages   = {127--190},
  year    = {1999},
  doi     = {10.1023/A:1010091220143}
}

@book{khalil2002nonlinear,
  author    = {Khalil, Hassan K.},
  title     = {Nonlinear Systems},
  edition   = {3rd},
  publisher = {Prentice Hall},
  year      = {2002}
}

@inproceedings{nguyen2016ecbf,
  author    = {Nguyen, Quan and Sreenath, Koushil},
  title     = {Exponential Control Barrier Functions for Enforcing High Relative-Degree
               Safety-Critical Constraints},
  booktitle = {American Control Conference (ACC)},
  year      = {2016},
}

@inproceedings{agrawal2017dcbf,
  author    = {Agrawal, Ayush and Sreenath, Koushil},
  title     = {Discrete Control Barrier Functions for Safety-Critical Control of
               Discrete Systems with Application to Bipedal Robot Navigation},
  booktitle = {Robotics: Science and Systems (RSS)},
  year      = {2017},
  doi       = {10.15607/RSS.2017.XIII.073}
}

@inproceedings{khazoom2022humanoidcbf,
  author    = {Khazoom, Charles and Gonz{\'a}lez-D{\'i}az, Daniel and
               Ding, Yanran and Kim, Sangbae},
  title     = {Humanoid Self-Collision Avoidance Using Whole-Body Control with
               Control Barrier Functions},
  booktitle = {IEEE-RAS International Conference on Humanoid Robots},
  year      = {2022},
}

@inproceedings{xiao2019hrdcbf,
  author    = {Xiao, Wei and Belta, Calin},
  title     = {Control Barrier Functions for Systems with High Relative Degree},
  booktitle = {IEEE Conference on Decision and Control (CDC)},
  year      = {2019},
}

@inproceedings{xu2015robustcbf,
  author    = {Xu, Xiangru and Tabuada, Paulo and Grizzle, Jessy W. and Ames, Aaron D.},
  title     = {Robustness of Control Barrier Functions for Safety Critical Control},
  booktitle = {IFAC Conference on Analysis and Design of Hybrid Systems (ADHS)},
  year      = {2015},
}

@inproceedings{yun2025siadaptation,
  author    = {Yun, Kai S. and Chen, Rui and Dunaway, Chase and Dolan, John M.
               and Liu, Changliu},
  title     = {Safe Control of Quadruped in Varying Dynamics via Safety Index Adaptation},
  booktitle = {IEEE International Conference on Robotics and Automation (ICRA)},
  year      = {2025},
  eprint    = {2409.09882},
  archivePrefix = {arXiv},
  doi       = {10.1109/ICRA55743.2025.11127362}
}

@inproceedings{zhao2021modelfree,
  author    = {Zhao, Weiye and He, Tairan and Liu, Changliu},
  title     = {Model-free Safe Control for Zero-Violation Reinforcement Learning},
  booktitle = {Conference on Robot Learning (CoRL)},
  year      = {2021},
}

@article{dawson2023safecert,
  author  = {Dawson, Charles and Gao, Sicun and Fan, Chuchu},
  title   = {Safe Control With Learned Certificates: A Survey of Neural {Lyapunov},
             Barrier, and Contraction Methods},
  journal = {IEEE Transactions on Robotics},
  year    = {2023},
  eprint  = {2202.11762},
  archivePrefix = {arXiv},
  doi     = {10.1109/TRO.2022.3232542}
}

@inproceedings{peng2025biped,
  author    = {Peng, Chengyang and Paredes, Victor and Castillo, Guillermo A.
               and Hereid, Ayonga},
  title     = {Real-Time Safe Bipedal Robot Navigation using Linear Discrete
               Control Barrier Functions},
  booktitle = {IEEE International Conference on Robotics and Automation (ICRA)},
  year      = {2025},
  eprint    = {2411.03619},
  archivePrefix = {arXiv},
  doi       = {10.1109/ICRA55743.2025.11128366}
}

@article{mitchell2005hj,
  author  = {Mitchell, Ian M. and Bayen, Alexandre M. and Tomlin, Claire J.},
  title   = {A Time-Dependent {Hamilton-Jacobi} Formulation of Reachable Sets for
             Continuous Dynamic Games},
  journal = {IEEE Transactions on Automatic Control},
  volume  = {50},
  number  = {7},
  pages   = {947--957},
  year    = {2005},
  doi     = {10.1109/TAC.2005.851439}
}

@inproceedings{prajna2004barrier,
  author    = {Prajna, Stephen and Jadbabaie, Ali},
  title     = {Safety Verification of Hybrid Systems Using Barrier Certificates},
  booktitle = {Hybrid Systems: Computation and Control (HSCC)},
  series    = {Lecture Notes in Computer Science},
  year      = {2004},
  doi       = {10.1007/978-3-540-24743-2_32}
}

@inproceedings{gurriet2018activeset,
  author    = {Gurriet, Thomas and Singletary, Andrew W. and Reher, Jake and
               Ciarletta, Laurent and Feron, Eric M. and Ames, Aaron D.},
  title     = {Towards a Framework for Realizable Safety Critical Control through
               Active Set Invariance},
  booktitle = {ACM/IEEE International Conference on Cyber-Physical Systems (ICCPS)},
  year      = {2018},
}

@inproceedings{ames2019cbfsurvey,
  author    = {Ames, Aaron D. and Coogan, Samuel and Egerstedt, Magnus and
               Notomista, Gennaro and Sreenath, Koushil and Tabuada, Paulo},
  title     = {Control Barrier Functions: Theory and Applications},
  booktitle = {European Control Conference (ECC)},
  pages     = {3420--3431},
  year      = {2019}
}

@misc{surgery,
  author       = {{UC San Diego Today}},
  title        = {Surgeons Use Teleoperated Humanoid Robots to Perform Live Surgery},
  howpublished = {\url{https://today.ucsd.edu/story/surgeons-use-teleoperated-humanoid-robots-to-perform-live-surgery-a-world-first}},
  year         = {2026},
}

@inproceedings{fioraldi2020afl++,
  title={AFL++: Combining incremental steps of fuzzing research},
  author={Fioraldi, Andrea and Maier, Dominik and Eissfeldt, Heiko and Heuse, Marc},
  booktitle={14th USENIX Workshop on Offensive Technologies (WOOT 20)},
  year={2020},
}

@misc{unitreefirmware,
  author       = {{Bin4ry}},
  title        = {UniTEABag: Unitree UPK Firmware Update Decryption and Encryption},
  howpublished = {\url{https://github.com/Bin4ry/UniTEABag}},
  year         = {2026},
}

@inproceedings{Chen2018Hawkeye,
  author    = {Hongxu Chen and Bihuan Chen and Yinxing Xue and
               Xiaofei Xie and Yang Liu and Yong Li and
               Xiaohong Wu},
  title     = {Hawkeye: Towards a Desired Directed Grey-Box Fuzzer},
  booktitle = {Proceedings of the 2018 ACM SIGSAC Conference on
               Computer and Communications Security (CCS)},
  pages     = {2095--2108},
  year      = {2018},
  publisher = {ACM},
}

@inproceedings{Bohme2017AFLGo,
  author    = {Marcel B{\"o}hme and Van{-}Thuan Pham and
               Manh{-}Dung Nguyen and Abhik Roychoudhury},
  title     = {Directed Greybox Fuzzing},
  booktitle = {Proceedings of the 2017 ACM SIGSAC Conference on
               Computer and Communications Security (CCS)},
  pages     = {2329--2344},
  year      = {2017},
  publisher = {ACM},
}

@inproceedings{Lyu2019MOPT,
  author    = {Chenyang Lyu and Shouling Ji and Chao Zhang and
               Yuwei Li and Wei{-}Han Lee and Yu Song and
               Raheem Beyah},
  title     = {{MOPT}: Optimized Mutation Scheduling for Fuzzers},
  booktitle = {28th USENIX Security Symposium (USENIX Security 19)},
  pages     = {1949--1966},
  year      = {2019},
  publisher = {USENIX Association}
}

@inproceedings{Chen2018Angora,
  author    = {Peng Chen and Hao Chen},
  title     = {Angora: Efficient Fuzzing by Principled Search},
  booktitle = {Proceedings of the 2018 IEEE Symposium on Security and Privacy (SP)},
  pages     = {711--725},
  year      = {2018},
  publisher = {IEEE},
}

@misc{cveble,
  author       = {{NIST}},
  title        = {National Vulnerability Database: {CVE-2025-35027}},
  howpublished = {\url{https://nvd.nist.gov/vuln/detail/CVE-2025-35027}},
  year         = {2025},
}

@misc{cvewifi,
  author       = {{NIST}},
  title        = {National Vulnerability Database: {CVE-2025-60017}},
  howpublished = {\url{https://nvd.nist.gov/vuln/detail/CVE-2025-60017}},
  year         = {2025},
}

@misc{cvedds,
  author       = {{NIST}},
  title        = {National Vulnerability Database: {CVE-2026-27509}},
  howpublished = {\url{https://nvd.nist.gov/vuln/detail/CVE-2026-27509}},
  year         = {2026},
}

@inproceedings{koscher2010car,
    author    = {Koscher, Karl and Czeskis, Alexei and Roesner, Franziska and Patel, Shwetak and Kohno, Tadayoshi and Checkoway, Stephen and McCoy, Damon and Kantor, Brian and Anderson, Danny and Shacham, Hovav and
    Savage, Stefan},
    title     = {Experimental Security Analysis of a Modern Automobile},
    booktitle = {IEEE Symposium on Security and Privacy},
    pages     = {447--462},
    year      = {2010},
  }

@inproceedings{checkoway2011car,
    author    = {Checkoway, Stephen and McCoy, Damon and Kantor, Brian and Anderson, Danny and Shacham, Hovav and Savage, Stefan and Koscher, Karl and Czeskis, Alexei and Roesner, Franziska and Kohno, Tadayoshi},
    title     = {Comprehensive Experimental Analyses of Automotive Attack Surfaces},
    booktitle = {USENIX Security Symposium},
    year      = {2011},
  }

@inproceedings{cao2019lidar,
    author    = {Cao, Yulong and Xiao, Chaowei and Cyr, Benjamin and Zhou, Yimeng and Park, Won and Rampazzi, Sara and Chen, Qi Alfred and Fu, Kevin and Mao, Z. Morley},
    title     = {Adversarial Sensor Attack on {LiDAR}-based Perception in Autonomous Driving},
    booktitle = {ACM SIGSAC Conference on Computer and Communications Security},
    pages     = {2267--2281},
    year      = {2019},
  }

@article{mayoral2018rsf,
    author        = {Mayoral Vilches, V{\'i}ctor and Alzola Kirschgens, Laura and Bilbao Calvo, Asier and Hern{\'a}ndez Cordero, Alejandro and Izquierdo Pis{\'o}n, Rodrigo and Mayoral Vilches, David and Mu{\~n}iz
    Rosas, Aday and Olalde Mendia, Gorka and Usategi San Juan, Lander and Zamalloa Ugarte, Irati and Gil-Uriarte, Endika and Tews, Erik and Peter, Andreas},
    title         = {Introducing the Robot Security Framework ({RSF}), a Standardized Methodology to Perform Security Assessments in Robotics},
    journal       = {arXiv preprint arXiv:1806.04042},
    year          = {2018},
  }

@INPROCEEDINGS{mayoral2019akerbeltz,
  author={Mayoral-Vilches, Víctor and Carbajo, Unai Ayucar and Gil-Uriarte, Endika},
  booktitle={2020 Fourth IEEE International Conference on Robotic Computing (IRC)},
  title={Industrial robot ransomware: Akerbeltz},
  year={2020},
  volume={},
  number={},
  pages={432-435},
}

@article{yaacoub2021robotics,
    author  = {Yaacoub, Jean-Paul A. and Noura, Hassan N. and Salman, Ola and Chehab, Ali},
    title   = {Robotics Cyber Security: Vulnerabilities, Attacks, Countermeasures, and Recommendations},
    journal = {International Journal of Information Security},
    volume  = {21},
    pages   = {115--158},
    year    = {2022},
  }

@inproceedings{sun2020lidar,
  author    = {Sun, Jiachen and Cao, Yulong and Chen, Qi Alfred and Mao, Z. Morley},
  title     = {Towards Robust {LiDAR}-based Perception in Autonomous Driving:
               General Black-box Adversarial Sensor Attack and Countermeasures},
  booktitle = {USENIX Security Symposium},
  year      = {2020}
}

@inproceedings{zhang2025,
author = {Zhang, Yan and Liu, Zihao and Zhu, Yi and Miao, Chenglin},
title = {Towards Real-Time Defense against Object-Based LiDAR Attacks in Autonomous Driving},
year = {2025},
isbn = {9798400715259},
publisher = {Association for Computing Machinery},
address = {New York, NY, USA},
booktitle = {Proceedings of the 2025 ACM SIGSAC Conference on Computer and Communications Security},
pages = {3825–3839},
numpages = {15},
location = {Taipei, Taiwan},
series = {CCS '25}
}

@inproceedings{shen2020drift,
  author    = {Shen, Junjie and Won, Jun Yeon and Chen, Zeyuan and Chen, Qi Alfred},
  title     = {Drift with Devil: Security of Multi-Sensor Fusion based Localization
               in High-Level Autonomous Driving under {GPS} Spoofing},
  booktitle = {USENIX Security Symposium},
  year      = {2020}
}

@inproceedings{cao2021invisible,
  author    = {Cao, Yulong and Wang, Ningfei and Xiao, Chaowei and Yang, Dawei and
               Fang, Jin and Yang, Ruigang and Chen, Qi Alfred and Liu, Mingyan and Li, Bo},
  title     = {Invisible for Both Camera and {LiDAR}: Security of Multi-Sensor
               Fusion based Perception in Autonomous Driving Under Physical-World Attacks},
  booktitle = {IEEE Symposium on Security and Privacy},
  year      = {2021}
}

@inproceedings{kim2021pgfuzz,
  author    = {Kim, Hyungsub and Ozmen, Muslum Ozgur and Bianchi, Antonio and
               Celik, Z. Berkay and Xu, Dongyan},
  title     = {{PGFUZZ}: Policy-Guided Fuzzing for Robotic Vehicles},
  booktitle = {Network and Distributed System Security Symposium},
  year      = {2021}
}

@inproceedings{muller2022drivetruth,
  author    = {Muller, Raymond and Man, Yanmao and Celik, Z. Berkay and Li, Ming and Gerdes, Ryan},
  title     = {{DriveTruth}: Automated Autonomous Driving Dataset Generation for Security Applications},
  booktitle = {International Workshop on Automotive and Autonomous Vehicle Security},
  year      = {2022}
}

@inproceedings{song2023acero,
  author    = {Song, Ruoyu and Ozmen, Muslum Ozgur and Kim, Hyungsub and
               Muller, Raymond and Celik, Z. Berkay and Bianchi, Antonio},
  title     = {Discovering Adversarial Driving Maneuvers against Autonomous Vehicles},
  booktitle = {USENIX Security Symposium},
  year      = {2023}
}

@inproceedings{kim2023patchverif,
  author    = {Kim, Hyungsub and Ozmen, Muslum Ozgur and Celik, Z. Berkay and
               Bianchi, Antonio and Xu, Dongyan},
  title     = {{PatchVerif}: Discovering Faulty Patches in Robotic Vehicles},
  booktitle = {USENIX Security Symposium},
  year      = {2023}
}

@inproceedings{yeke2025raven,
  author    = {Yeke, Doguhan and Pant, Kartik A. and Ozmen, Muslum Ozgur and
               Kim, Hyungsub and Goppert, James M. and Hwang, Inseok and
               Bianchi, Antonio and Celik, Z. Berkay},
  title     = {Automated Discovery of Semantic Attacks in Multi-Robot Navigation Systems},
  booktitle = {USENIX Security Symposium},
  year      = {2025}
}

@article{mayoral2025humanoid,
  author        = {Mayoral-Vilches, V{\'i}ctor},
  title         = {The Cybersecurity of a Humanoid Robot: An Early Study on the
                   Cybersecurity of Humanoid Robots via the {Unitree G1}},
  journal       = {arXiv preprint arXiv:2509.14096},
  year          = {2025},
}

@article{mayoral2025humanoidattackvector,
  author        = {Mayoral-Vilches, V{\'i}ctor},
  title         = {Cybersecurity {AI}: Humanoid Robots as Attack Vectors},
  journal       = {arXiv preprint arXiv:2509.14139},
  year          = {2025},
}

\clearpage
\appendix
\section*{Ethical Considerations}

As \safecontrols for humanoid robots remain an emerging technology, the \safecontrols evaluated in this work are still undergoing industry validation and have not yet been productized or integrated into the default configurations used in humanoid robot manufacturing.
The goal of this work is to proactively discover critical and adversarial safety issues before any humanoids being deployed.
We will work with the respective authors to identify downstream users who may have deployed the affected \safecontrols and to inform them of the vulnerabilities identified in this work.

\section{Clearance Derivation}
\label{app:fixed_rate_clearance}

This appendix formalizes the collision possibility by deriving the
closed-form expression for the clearance $d(t)$ between a humanoid
collision volume and an obstacle.  Let $t=0$ denote the state
$\mathbf{x}_0$ at which the relevant safety constraint engages, and write
$d(t)\equiv d(\mathbf{x}(t))$.  Following the notation in
Section~\ref{sec:attack_surface}, assume over the local interval of
interest that the achieved value of the left-hand side of
Equation~\ref{eq:clearance_ode_general} is constant:
\begin{equation}
\dot{d}(t)
+
k_1\ddot{d}(t)
+
k_2\dddot{d}(t)
+
\cdots
+
k_{r-1}d^{(r)}(t)
=
\actual,
\qquad
\actual \geq \rate.
\label{eq:appendix_clearance_ode_constant_rate}
\end{equation}
This assumption is exact over an interval where the achieved escape rate
is constant; more generally, it provides a local model of the clearance
response until the clearance either recovers or reaches zero.

Define
\begin{equation}
z(t)
=
d(t)+k_1\dot{d}(t)+\cdots+k_{r-1}d^{(r-1)}(t).
\end{equation}
Equation~\ref{eq:appendix_clearance_ode_constant_rate} gives
$\dot{z}(t)=\actual$. Moreover,
Equation~\ref{eq:distance_safety_index} gives
$z(0)=d_{\min}-\phi(\mathbf{x}_0)$. Therefore,
\begin{equation}
d(t)+k_1\dot{d}(t)+\cdots+k_{r-1}d^{(r-1)}(t)
=
d_{\min}-\phi(\mathbf{x}_0)
+\actual t.
\label{eq:integrated_clearance_ode}
\end{equation}

Let
\begin{equation}
p(s)=1+k_1s+k_2s^2+\cdots+k_{r-1}s^{r-1}.
\label{eq:index_characteristic}
\end{equation}
Assume that every root of $p(s)$ has strictly negative real part. The
solution of
Equation~\ref{eq:integrated_clearance_ode} can then be written as
\begin{equation}
d(t)
=
A+\actual t+h(t),
\label{eq:clearance_general_solution}
\end{equation}
where
\begin{equation}
A
=
d_{\min}-\phi(\mathbf{x}_0)
-k_1\actual,
\label{eq:clearance_floor}
\end{equation}
and $h(t)$ is the homogeneous response needed to match the derivatives
of $d$ at engagement. Its coefficients also depend on
$\actual$, and it decays to zero under the stability
assumption. For distinct roots $s_1,\ldots,s_{r-1}$,
$h(t)=\sum_{j=1}^{r-1}c_j e^{s_jt}$; repeated roots produce the usual
polynomial--exponential terms.

The three terms in Equation~\ref{eq:clearance_general_solution} expose
the relevant behavior. The constant $A$ combines the engagement state
with a rate-dependent offset, the linear term captures recovery induced
by the achieved escape rate, and $h(t)$ captures the inherited approach
dynamics. A positive
$\actual$ therefore does not preclude an initial
decrease in $d(t)$.

To obtain a sufficient collision condition, first consider the
zero-rate solution, denoted by $d_0(t)$. From
Equations~\ref{eq:clearance_general_solution} and
\ref{eq:clearance_floor},
\begin{equation}
\lim_{t\rightarrow\infty}d_0(t)
=
d_{\min}-\phi(\mathbf{x}_0).
\end{equation}
Hence, if
\begin{equation}
\phi(\mathbf{x}_0)>d_{\min},
\label{eq:under_escape_condition}
\end{equation}
then the limiting clearance is negative. Even when $d(0)>0$, continuity
implies that $d_0(T)<0$ for some finite $T>0$. Solutions of the linear
ODE in Equation~\ref{eq:appendix_clearance_ode_constant_rate} depend continuously
on $\actual$. There consequently exists an
$\epsilon>0$ such that
\begin{equation}
0<\actual<\epsilon
\quad\Longrightarrow\quad
d(T)<0.
\end{equation}
Thus, for the same engagement state, a sufficiently small positive
achieved escape rate can satisfy the safety-index derivative constraint
throughout the fixed-rate interval while the physical clearance crosses
zero.

\end{document}